\documentclass[lettersize,journal]{IEEEtran}
\usepackage{amsmath,amsfonts}
\usepackage{algorithmic}
\usepackage{algorithm}
\usepackage{array}
\usepackage[caption=false,font=normalsize,labelfont=sf,textfont=sf]{subfig}
\usepackage{textcomp}
\usepackage{stfloats}
\usepackage{url}
\usepackage{verbatim}
\usepackage{graphicx}
\usepackage{cite}
\usepackage[colorlinks=true, allcolors=blue]{hyperref}

\usepackage{cancel}
\usepackage{color}
\usepackage{ulem}    % for removed text (strikethrough)

\begin{document}

\title{Generative Diffusion Model for Seismic Imaging Improvement of Sparsely Acquired Data and Uncertainty Quantification}

\author{Xingchen Shi, Shijun Cheng, Weijian Mao, and Wei Ouyang
\thanks{Manuscript submitted July 31, 2024. 
This work was supported by the National Natural Science Foundation of China under Grants 42204115 and 42130808.
\it{(Corresponding author: Shijun~Cheng)}}
\thanks{Xingchen Shi, Weijian Mao and Wei Ouyang are with the Research Center for Computational and Exploration Geophysics, State Key Laboratory of Geodesy and Earth's Dynamics, Innovation Academy for Precision Measurement Science and Technology, Chinese Academy of Sciences, Wuhan 430077, China (e-mail: shixingchen@apm.ac.cn; wjmao@whigg.ac.cn; ouyangwei@whigg.ac.cn).}
\thanks{Shijun Cheng is with the Division of Physical Science and Engineering, King Abdullah University of Science and Technology, Thuwal 23955-6900, Saudi Arabia (e-mail: sjcheng.academic@gmail.com).}
}

% The paper headers
\markboth{SUBMIT TO IEEE TRANSACTIONS ON GEOSCIENCE AND REMOTE SENSING}%
{Shell \MakeLowercase{\textit{et al.}}: A Sample Article Using IEEEtran.cls for IEEE Journals}

\maketitle

\begin{abstract}
Seismic imaging from sparsely acquired data faces challenges such as low image quality, discontinuities, and migration swing artifacts. Existing convolutional neural network (CNN)-based methods struggle with complex feature distributions and cannot effectively assess uncertainty, making it hard to evaluate the reliability of their processed results. To address these issues, we propose a new method using a generative diffusion model (GDM). Here, in the training phase, we use the imaging results from sparse data as conditional input, combined with noisy versions of dense data imaging results, for the network to predict the added noise. After training, the network can predict the imaging results for test images from sparse data acquisition, using the generative process with conditional control. This GDM not only improves image quality and removes artifacts caused by sparse data, but also naturally evaluates uncertainty by leveraging the probabilistic nature of the GDM. To overcome the decline in generation quality and the memory burden of large-scale images, we develop a patch fusion strategy that effectively addresses these issues. Synthetic and field data examples demonstrate that our method significantly enhances imaging quality and provides effective uncertainty quantification.

\end{abstract}

\section{Introduction}
\IEEEPARstart{S}{eismic} high-resolution imaging is crucial for accurately mapping subsurface structures and characterizing complex geological targets. For high-resolution imaging of the subsurface, the ideal scenario involving dense acquisition geometries, including both sources and receivers, is required in a seismic survey \cite{vermeer19983}. However, practical limitations sometimes result in sparse data, where spatial sampling for high-resolution imaging is inadequate.
For example, in deep-water ocean bottom node (OBN) acquisition surveys, due to cost considerations, the distribution of receivers is very sparse, negatively impacting high-resolution imaging. Such a sparsity can lead to aliasing, artifacts, and a general degradation of image quality, compromising the reliability of subsequent interpretations and analyses.
Moreover, the insufficient illumination caused by sparse sampling produces acquisition footprints in the imaging profile, which further reduces the imaging resolution and alters the amplitude variation with offset (AVO) response characteristics. In such cases, the quality of the results produced by standard migration methods is poor.

Common strategies to overcome the limitation of sparse data imaging include regularization, interpolation, and incorporation of a priori information to constrain the imaging process. 
In practice, trace interpolation and regularization are frequently employed to fill in gaps in sparsely acquired data by interpolating missing traces and regularizing the data to ensure smoothness and continuity. Techniques such as the Fourier transform\cite{naghizadeh2011seismic}, the Radon transform\cite{chen_fast_2022}, and the seislet transform\cite{fomel_seislet_2010}, as well as other sparse transforms, are commonly used for this purpose. 
According to Baykulov and Gajewski\cite{baykulov_prestack_2009}, the common-reflection-surface stacking method can also be used to regularize and interpolate sparse low-fold seismic data.
However, these methods can be challenging to implement with complex geological structures and computationally intensive when dealing with large datasets.
Traditional seismic imaging techniques have long recognized the challenges associated with sparse data. To mitigate these issues, several strategies have been proposed, including limiting the migration operator to regions near the specular reflection point. For example, Hu and Stoffa\cite{hu2009slowness} introduced a technique known as slowness-driven Gaussian-beam prestack depth migration. This method integrates prestack instantaneous slowness information into the imaging condition, effectively reducing migration swing artifacts and enhancing the image quality of low-fold seismic reflection data or data with acquisition geometry limitations.
In addition, least-squares migration (LSM) \cite{nemeth_least-squares_1999} can yield high-quality imaging results in scenarios with limited recording aperture, coarse sampling, and suboptimal acquisition geometries, but it faces limitations due to computational demands and the quality of seismic data.
Point spread functions (PSFs) can also be used to address the challenge of high-resolution imaging of sparsely sampled data \cite{lecomte_resolution_2008,fletcher_least-squares_2016}. However, methods based on PSFs are sensitive to noise and may not perform well with highly sparse data.
% However, those conventional methods face limitations due to computational demands and the quality of seismic data.

% \textcolor{red}{Deep Learning and Its Limitations:} 
Recently, the advancement of deep learning (DL) has offered a new avenue for various seismic processing tasks \cite{wu2019faultseg3d, yu2021deep, zhang2021deep, mousavi2022deep, harsuko2022storseismic, cheng2023effective, cheng2024meta, cheng2024self, mousavi2024applications}. DL models, such as convolutional neural networks (CNNs), have shown promise in image enhancement and noise suppression \cite{li2022deep, yu2023enhancing}. Significant attention has been devoted to reducing computational burdens and data acquisition costs. For example, Zhang et al. \cite{zhang2022deep} used a CNN architecture to present the optimal reflection image for a single-shot recording from the standard reverse-time migration (RTM) image and, thus, enhance the computational efficiency of prestack least-squares RTM (LSRTM). Torres and Sacchi \cite{torres2022least} used an iterative DL framework with a residual CNN and projected gradient descent to improve LSRTM. This work addressed the iteration inefficiencies and regularization challenges of LSRTM and achieved high-resolution reflectivity updates with fewer iterations. On the other hand, some researches focus on removing noise from the imaging results of sparse data and improving continuity. Picetti et al. \cite{picetti2019seismic} employed a GAN to recover an image obtained through RTM of a dense acquisition geometry from the migration result of a very coarse acquisition geometry. Cheng et al. \cite{cheng2023elastic} used a multi-scale CNN to address the issue of high-resolution imaging of sparse 4C OBN data. Dong et al. \cite{dong2024can} constructed a CNN with a self-guided attention network architecture to transform sparse-shot images into dense-shot images, effectively improving the quality of sparse-shot images to closely resemble dense-shot images. 

Although these DL methods are capable of providing impressive results, they fail to demonstrate the reliability of these results. This limitation is not unique to these methods, but is a common challenge among many DL approaches for imaging enhancement. Actually, in the context of enhancing imaging results derived from sparse data, uncertainty quantification plays a critical role. Sparse data commonly introduces higher levels of uncertainty due to the limited illumination, which can result in less reliable imaging products. Quantifying this uncertainty allows us to assess the confidence in the processed results, identify regions with high potential errors, and finally improve decision-making processes in applications relying on these imaging results. This step is essential to distinguish between areas where the predictions are dependable and areas where further investigation or alternative approaches might be necessary.

Recently, generative diffusion models (GDMs) have emerged as a powerful framework in the field of machine learning, leveraging a step-by-step denoising approach to synthesize high-quality samples from random noise\cite{ho2020denoising, song2020denoising, nichol2021improved, ho2022classifier}. These models operate by gradually diffusing the data distribution into Gaussian noise through a series of transformations and then reversing this process to reconstruct the original data. This method has garnered significant attention due to its ability to generate high-fidelity samples and its robustness in modeling complex data distributions.
Moreover, GDMs manage uncertainty more effectively by generating various samples according to the probability distribution, which is essential for applications needing insight into the spectrum of potential results. 

In the field of exploration seismology, some researchers have employed GDMs to perform some seismic processing tasks. For example, Durall et al. \cite{durall2023deep} used GDMs to address seismic data processing issues, such as demultiple, denoising, and interpolation, demonstrating their effectiveness over traditional methods through experiments on synthetic and field data. Wang et al. \cite{wang2023prior} pretrained the diffusion model on prior velocity distributions and integrated it with full-waveform inversion during sampling. As a result, this approach achieves high-resolution subsurface models even with sparse or noisy data. Wei et al. \cite{wei2023seismic} introduced a denoising diffusion implicit model (DDIM) with resampling to interpolate missing seismic data. Zhang et al. \cite{zhang2024conditional} proposed using a conditional denoising diffusion probabilistic model (DDPM) to effectively separate seismic diffractions from full-wavefield data for high-resolution imaging of small-scale geological targets. Li et al. \cite{li2024conditional} also introduced a conditional DDPM to efficiently attenuate ground-roll noise in land seismic surveys while preserving valuable reflection events. Zhang et al. \cite{zhang2024seisresodiff} proposed SeisResoDiff, a scheme using the DDPM to enhance the resolution of post-stack seismic data. Wang et al. \cite{wang2024self} developed a self-supervised diffusion model for the reconstruction of 3D seismic data, addressing the limitations of previous methods that rely on CNN and require extensive paired data. While these works have demonstrated the potential of GDMs in the field of seismic processing, most of them focus mainly on pre-stack or post-stack seismic data processing, with relatively less attention given to imaging enhancement tasks. Meanwhile, more studies do not effectively utilize the probabilistic nature of GDMs for the uncertainty analysis of processing results.

In this paper, we propose a novel application of GDMs designed to enhance seismic imaging of sparsely acquired seismic data. To adapt to our task, we use the imaging results from sparse data as a condition during the network training process, along with noisy versions of the imaging results from dense data as the network input. The loss required for network optimization is derived from the mean squared error (MSE) between the network output and the noise added to the dense data. During the generation phase, the network iteratively denoises the input, which combines randomly sampled noise and imaging results from sparse data, ultimately producing our processed results. In addition, we develop a patch fusion strategy to address the limitations of GDMs, which are trained on small-sized data samples and exhibit reduced performance in generating large-scale imaging results. To provide a comprehensive uncertainty quantification of our processed results, we generate multiple imaging results conditioned on the same image from sparse data and measure the variability among these results. 
    
The contributions of this paper are summarized as follows:
\begin{itemize}
   \item We introduce a new paradigm for seismic imaging enhancement that leverages the strengths of GDMs. 
   \item We consider the probabilistic nature of GDMs to quantify the uncertainty of our processing results and, thus, improve decision-making processes.
   \item We develop a patch fusion strategy for the GDM sampling process that effectively reduces memory burden and addresses limitations in the generation of large-scale imaging results.
   \item We validate our approach on synthetic and field seismic datasets, showcasing its effectiveness in improving imaging quality and providing robust uncertainty estimates. 
\end{itemize}

The rest of the paper is structured as follows. First, we introduce the GDMs and present our modifications to enhance the seismic imaging results derived from the sparse data acquisition. Second, we illustrate how to implement the patch fusion strategy in the generation process and perform uncertainty quantification, a critical component of our approach. Third, we detail the network architecture used in this paper and also describe the network training configurations and the preparation of training data sets. Fourth, we share our test results to highlight the effectiveness of our method. Moreover, we analyze some key components of our approach in the Discussion section. Finally, we conclude our work.

\section{Method}
\subsection{Generative Diffusion Models} 
\begin{figure}[!t]
\centering
\includegraphics[width=3in]{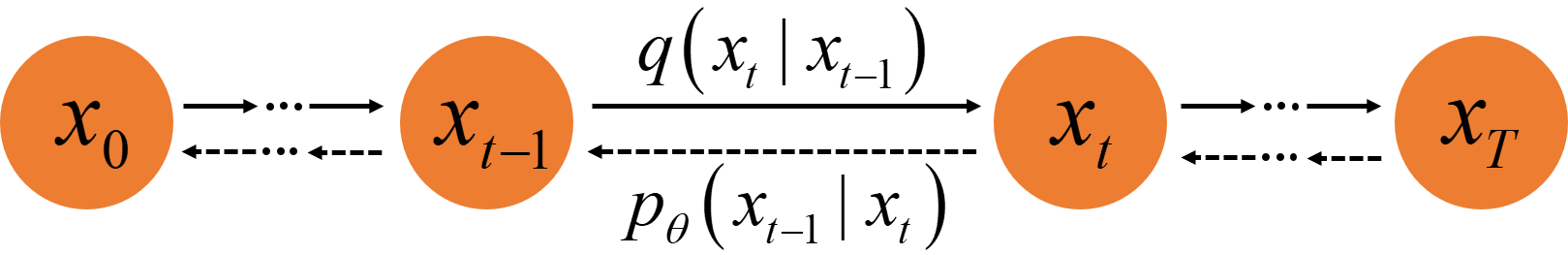}
\caption{An illustration of DDPM. In the forward diffusion process, $x_0 \rightarrow x_{T}$, the data is gradually transformed into noise through a series of intermediate latent states $x_{t}$, followed by $q(x_t|x_{t-1})$. The reverse process $x_{T} \rightarrow x_0$ involves denoising the latent states back to the data, guided by the learned probability distributions $p_\theta(x_{t-1}|x_t)$.}
\label{fig1}
\end{figure}

A prominent example of GDMs is the DDPM. The DDPM framework, which is illustrated in Fig. \ref{fig1}, involves two stages: the forward diffusion process and the reverse denoising process. In the forward diffusion process, the model progressively adds noise to the original data through a multi-step Markov chain. For a given data sample $x_0$, this process generates a sequence of increasingly noisy samples $x_1$, $x_2$, $\ldots$, $x_T$. Each step in the forward process is defined as \cite{ho2020denoising}:
\begin{equation}\label{eq1}
q(x_t | x_{t-1}) = \mathcal{N}(x_t; \sqrt{\alpha_t} x_{t-1}, (1 - \alpha_t) \mathbf{I}).
\end{equation}
In (\ref{eq1}), $x_t$ is the noisy version of the data sample at step $t$, $\alpha_t$ is a parameter that controls the rate of noise addition, $\mathcal{N}$ denotes a Gaussian distribution, and $\mathbf{I}$ is the identity matrix, ensuring isotropic Gaussian noise. Specifically, $\alpha_t = 1 - \beta_t$, where $\beta_t$ represents a small positive constant that dictates the variance of the noise added at each step. The sequence effectively diffuses the original data into a pure Gaussian noise distribution over $T$ steps.

To make the process more clear, we denote the cumulative product of $\alpha_t$ up to step $t$ as follows:
\begin{equation}\label{eq2}
\bar{\alpha}_t = \prod_{i=1}^t \alpha_i.
\end{equation}
Using $\bar{\alpha}_t$, we can express $x_t$ in a more explicit form:
\begin{equation}\label{eq3}
x_t = \sqrt{\bar{\alpha}_t} x_0 + \sqrt{1 - \bar{\alpha}_t} \epsilon,
\end{equation}
where $\epsilon \sim \mathcal{N}(0, \mathbf{I})$ represents the Gaussian noise added at each step. By reformulating $x_t$ in this way, we can clearly separate the original data component from the noise component. This separation is crucial for training the denoising network. The objective of the network is to predict the noise component $\epsilon$ given the noisy data $x_t$.

To achieve this, the training process involves minimizing the difference between the predicted noise and the actual noise added during the forward diffusion process. This is achieved by optimizing the network parameters $\theta$ to minimize the following loss function \cite{ho2020denoising}:
\begin{equation}\label{eq4}
L(\theta) = \mathbb{E}_{t, x_0, \epsilon} \left[ \| \epsilon - \epsilon_\theta(x_t, t) \|^2 \right],
\end{equation}
where $t$ is sampled uniformly from $\{1, 2, ..., T\}$, $x_0$ is the original data sample, $\epsilon$ is the noise sampled from $\mathcal{N}(0, \mathbf{I})$, $x_t$ is given by (\ref{eq3}), and $\epsilon_\theta(x_t, t)$ is the noise predicted by the denoising network parameterized by $\theta$. By minimizing this loss function, the network learns to accurately predict the noise component at each step, enabling the effective removal of noise during the reverse process.

With the denoising network trained to predict the noise accurately, the reverse process can be performed. Starting from the completely noisy sample $x_T$, the trained denoising network is employed to progressively reduce the noise, thereby reconstructing the original data $x_0$. This reverse process follows a Markov chain, described by \cite{ho2020denoising}:
\begin{equation}\label{eq5}
p_\theta(x_{t-1} | x_t) = \mathcal{N}(x_{t-1}; \mu_\theta(x_t, t), \Sigma_\theta(x_t, t)),
\end{equation}
where $\mu_\theta(x_t, t)$ and and $\Sigma_\theta(x_t, t)$ are the mean and the variance predicted by the trained network parameterized by $\theta$, respectively.

Specifically, the reverse denoising process is implemented as follows:
\begin{enumerate}
    \item \textbf{Initialization}: Begin with the noisy sample $x_T$, which is drawn from a Gaussian distribution $\mathcal{N}(0, \mathbf{I})$.

    \item \textbf{Step-wise Denoising}:
    \begin{enumerate}
        \item For each time step $t$ from $T$ to 1, perform the following sub-steps:
        \begin{enumerate}
            \item If $t > 1$, sample $z$ from $\mathcal{N}(0, \mathbf{I})$. Otherwise, set $z = 0$.
            \item Use the trained denoising network to predict the noise component $\epsilon_\theta(x_t, t)$.
            \item Compute the mean $hat{x}_{t-1}$ of the posterior distribution as:
            \[
            {x}_{t-1} = \frac{1}{\sqrt{\alpha_t}} \left( x_t - \frac{1 - \alpha_t}{\sqrt{1 - \bar{\alpha}_t}} \epsilon_\theta(x_t, t) \right) + \sigma_t z,
            \]
            where $\sigma_t$ is the standard deviation of the noise to be added back, ensuring the correct variance for the distribution.
        \end{enumerate}
    \end{enumerate}
    
    \item \textbf{Update}: Update $x_t$ to $x_{t-1}$ for the next iteration.

    \item \textbf{Final Output}: After completing all steps, the final output $x_0$ is obtained, which should closely approximate the original data sample.
\end{enumerate}
We can see that, the denoising network learns to map noisy samples back to less noisy ones, effectively inverting the diffusion process. 

Although DDPM can generate high-quality results, it follows a Markov chain process, requiring a large number of sampling steps (typically thousands of steps) to achieve good generative performance. This means that generating a single product involves using the trained network to make thousands of predictions, which is time-consuming. In contrast, DDIM \cite{song2020denoising} introduces a deterministic denoising process that significantly reduces the number of sampling steps (down to a few dozen steps), making the generation process faster. Essentially, DDPM and DDIM maintain the same forward diffusion process. That is, they can share the same trained denoising network. The differences between them arise from the reverse process. Unlike the stochastic nature of DDPM's reverse process, DDIM employs a deterministic approach, which simplifies the reverse diffusion into a more efficient procedure.

In DDIM, the reverse process can be formulated as \cite{song2020denoising}:
\begin{equation}\label{eq6}
\begin{aligned}
x_{t-1} &= \sqrt{\bar{\alpha}_{t-1}} \left( \frac{x_t - \sqrt{1 - \bar{\alpha}_t} \epsilon_\theta(x_t, t)}{\sqrt{\bar{\alpha}_t}} \right) \\
&\quad + \sqrt{1 - \bar{\alpha}_{t-1} - \sigma_t^2} \epsilon_\theta(x_t, t) + \sigma_t z.
\end{aligned}
\end{equation}
Here, when the term $\sigma_t$ is equal to $\sqrt{(1 - \bar{\alpha}_{t-1})/(1 - \bar{\alpha}_{t})}\sqrt{(1 - \bar{\alpha}_{t}/\bar{\alpha}_{t-1}}$ for all step $t$, the forward process becomes Markovian and, thus, the generative process becomes a DDPM \cite{song2020denoising}. In particular, if $\sigma_t$ is equal to zero at all time steps, then the forward process will be deterministic.

\begin{figure*}[!t]
\centering
\includegraphics[width=6in]{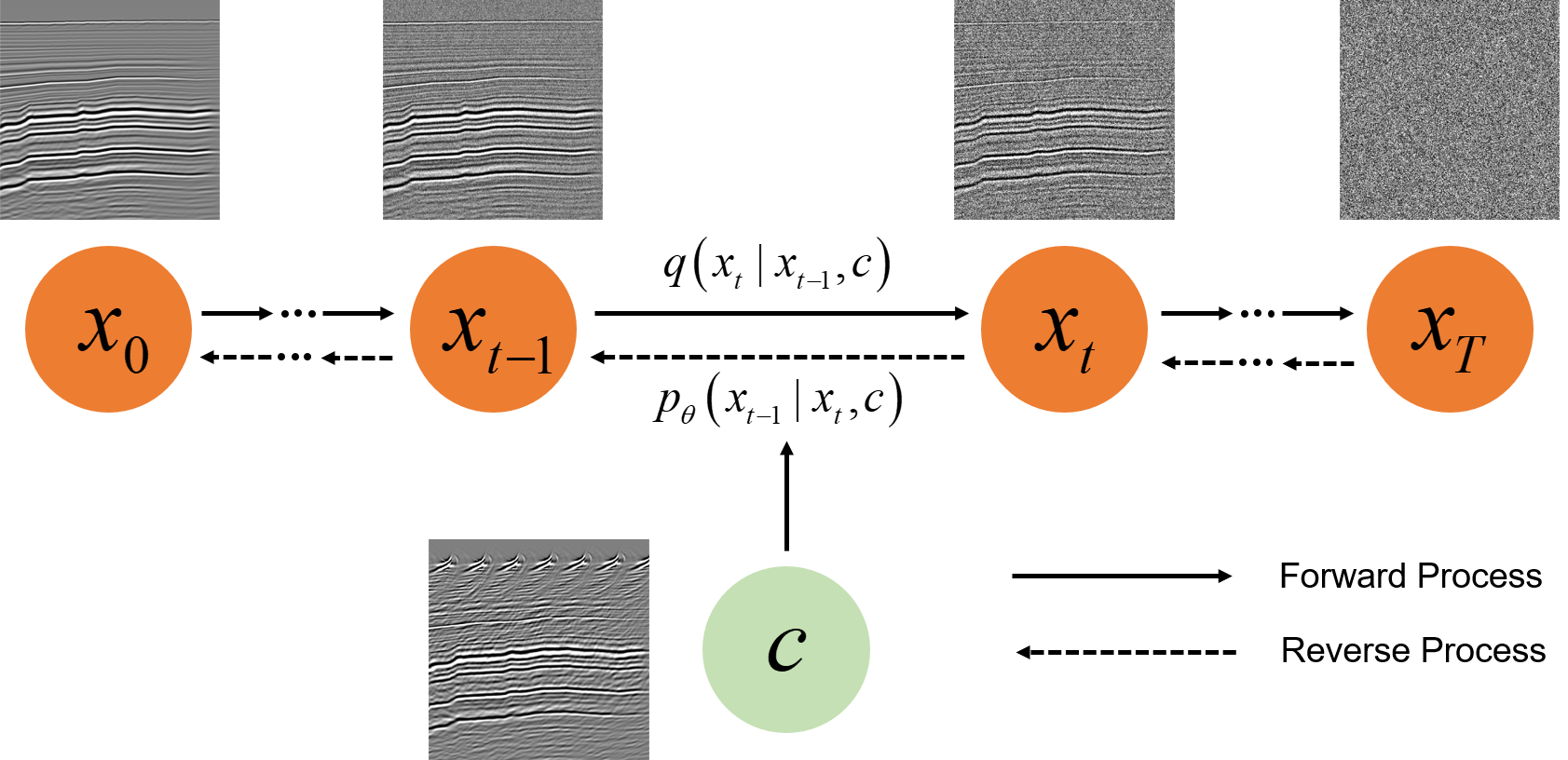}
\caption{An illustration of our adaptation using DDIM for seismic imaging enhancement. The forward process $x_0 \rightarrow x_T$ transforms the densely sampled seismic data $x_0$ into noise through intermediate latent states $x_t$. The reverse process $x_T \rightarrow x_0$ reconstructs the densely sampled seismic data from the noisy latent states using the learned conditional probability distributions $p_\theta(x_{t-1}|x_t, c)$, where $c$ represents the imaging results from sparsely sampled data.}
\label{fig2}
\end{figure*}

\subsection{Our Adaption}
The theoretical foundations of DDPM and DDIM provide a robust framework for leveraging GPMs in various domains. Here, we illustrate our adaptation of DDIM for enhancing seismic imaging results obtained from sparse data acquisition. 

To adapt DDIM for our specific task, we introduce the sparse data imaging result as a condition. This conditional information is embedded directly into the network by concatenating it with the noisy versions of the dense data imaging results. The network then uses this combined input to predict the added noise. 

Fig. \ref{fig2} illustrates our adaptation. Starting from the left, $x_0$ represents the dense data imaging result. As we move to the right, $x_{t-1}$ and $x_{t}$ depict intermediate noisy versions generated during the forward diffusion process. $x_{T}$ represents the final noisy version before refinement. The sparse data imaging result, denoted as $c$, is concatenated with the noisy versions at each step to condition the network's predictions. In the reverse process, the network processes this concatenated input to predict the added noise, ultimately refining $x_{T}$ to produce a high-quality imaging output. 

To summarize, the whole process is as follows: \\
\textbf{Input Preparation}:
\begin{enumerate}
\item Generate noisy versions of the dense data imaging results.
\item Concatenate these noisy versions with the sparse data imaging results.
\end{enumerate}
\textbf{Network Training}:
\begin{enumerate}
    \item Use a neural network to process the concatenated inputs.
    \item The network is trained to predict the noise added to the dense data imaging results.
\end{enumerate}
\textbf{Diffusion Process}:
\begin{enumerate}
    \item Utilize DDIM to iteratively refine the noisy inputs, conditioned on the sparse data imaging results, towards the final high-quality imaging output.
\end{enumerate}
By doing so, the network learns to leverage the structural information present in the images from sparse data while removing the noise to produce high-quality seismic images.

\subsection{Patch Fusion} 
\begin{figure}[!t]
\centering
\includegraphics[width=3in]{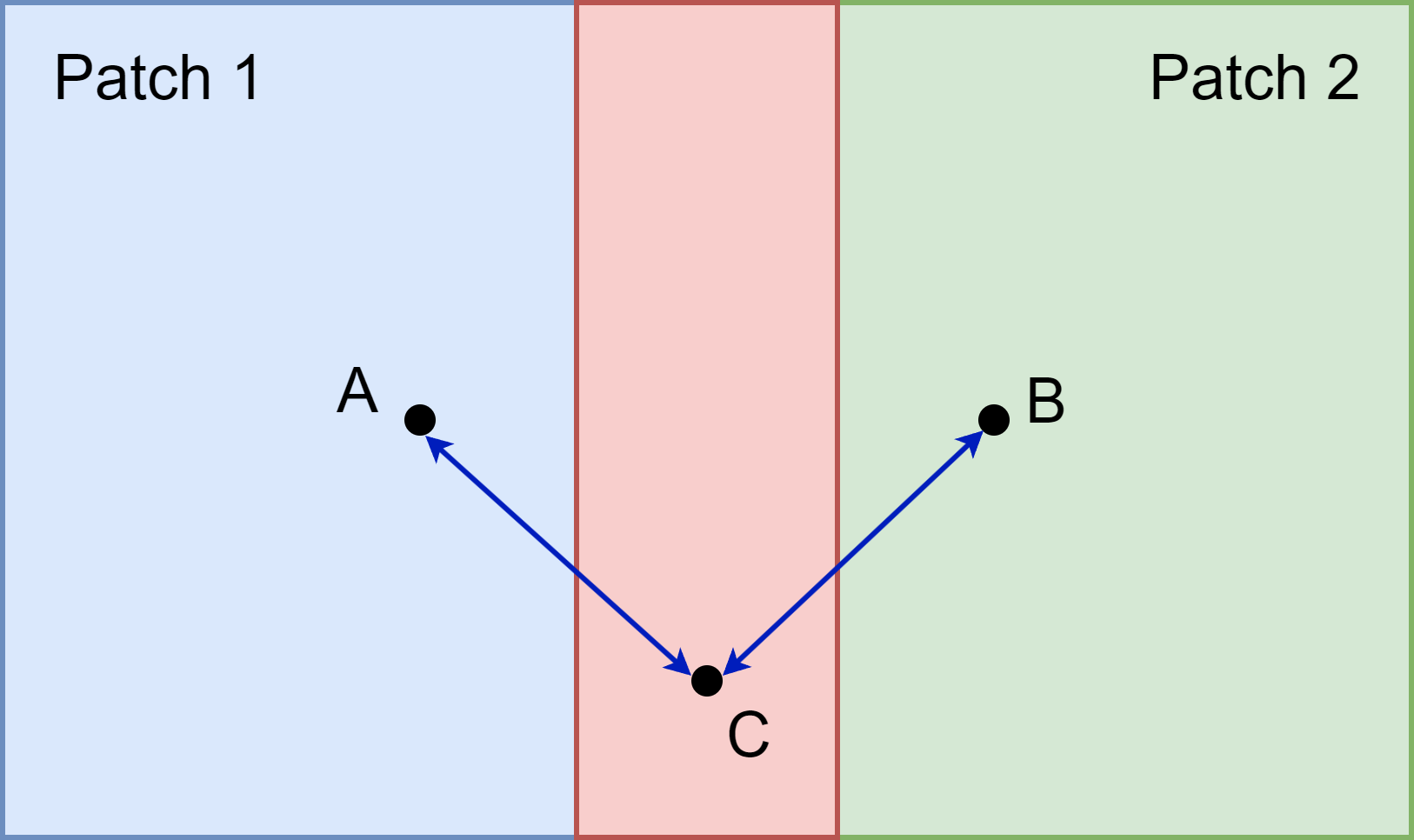}
\caption{An illustration of patch fusion. Points A and B are the center points of Patches 1 and 2, respectively. The pink area represents the overlapping region of the two patches. Point C is located at the overlapping region.}
\label{fig3}
\end{figure}

Due to the high memory consumption of GDMs, predicting large seismic images using a trained network, such as during the generation process, often exceeds GPU memory limits. To avoid excessive memory and time consumption during training, we typically use small patches, such as $128\times128$, for training. However, applying a network trained on small patches to predict large-scale seismic images leads to significant quality degradation in the prediction results. 

To address these challenges, we here propose a patch fusion method for processing large-scale imaging results, which is illustrated in Fig. \ref{fig3}. Specifically, we first divide the large seismic imaging results into smaller patches (e.g., Patches 1 and 2) with overlapping regions (e.g., pink area). Then, each patch is predicted separately, and the results are finally fused using a Gaussian weight to handle the overlaps. That is, the final prediction value for each grid point (e.g., point C) within the overlapping region is obtained by the weighted average of the predictions from each patch covering that area. The weighting coefficients for each patch are related to the distance between the grid point (e.g., point C) and the center of each patch (e.g., points A and B). This method ensures that the transition between patches is seamless, thus maintaining the quality of the final imaging results. 

Taking a example, given a large seismic image $I$ of size $H \times W $, we will perform the patch fusion method as follows: 
\begin{enumerate}
    \item \textbf{Split to patches}: We divide the image into smaller patches of size $p \times p$ with a step size $s$ such that each patch overlaps with its neighboring patches. Let $P_{i,j}$ represent the patch extracted from position $(i, j)$, where $i$ and $j$ vary from 0 to $H - p$ and $W - p$ respectively, with a step size of $s$.
    
    \item \textbf{Gaussian Weight Calculation}: To ensure smooth transitions between patches, we apply a Gaussian weight to each patch. The Gaussian kernel \( G(x, y) \) is defined as:
    \[
       G(x, y) = \exp \left( -\frac{x^2 + y^2}{2\sigma^2} \right),
    \]
    where $x$ and $y$ are the coordinates relative to the center of the patch, and $\sigma$ is set according to your situation.
    
    \item \textbf{Initialize Accumulated Prediction and Weight Matrices}: We initialize two zero matrices, $A$ and $W$, to store the accumulated predictions and weights, respectively. These matrices are of the same size as the input image $I$:
    \[
        A = \mathbf{0}_{H \times W},
    \]
    \[
        W = \mathbf{0}_{H \times W}.
    \]

    \item \textbf{Patch Prediction, Accumulate Predictions and Weights}: For each patch $P_{i,j}$, we use the DDIM sampling method to generate the predicted patch $\hat{P}_{i,j}$:
    \[ 
       \hat{P}_{i,j} = \text{DDIM}(P_{i,j}, \epsilon),
    \]
    where $\epsilon$ is the noise sampled from a Gaussian distribution $\mathcal{N}(0, \mathbf{I})$. Then, we add the weighted predicted patch to the accumulated matrices:
    \[
        A[i:i+p, j:j+p] += G(x, y) \odot \hat{P}_{i,j},
    \]
    \[
        W[i:i+p, j:j+p] += G(x, y),
    \]
    where $\odot$ denotes element-wise multiplication.

    \item \textbf{Final Image Reconstruction}: After processing all patches, the final reconstructed image \( \hat{I} \) is obtained by averaging the overlapping regions using the accumulated weights:
    \[
        \hat{I}(x, y) = \frac{A(x, y)}{W(x, y)},
    \]
    where $\hat{I}(x, y)$ is the final pixel value at position $(x, y)$.
\end{enumerate}
  
By adopting this patch fusion strategy, we effectively address the memory constraints and also ensure the generation quality of the network for large-scale imaging results.

\begin{figure*}[!t]
\centering
\includegraphics[width=7in]{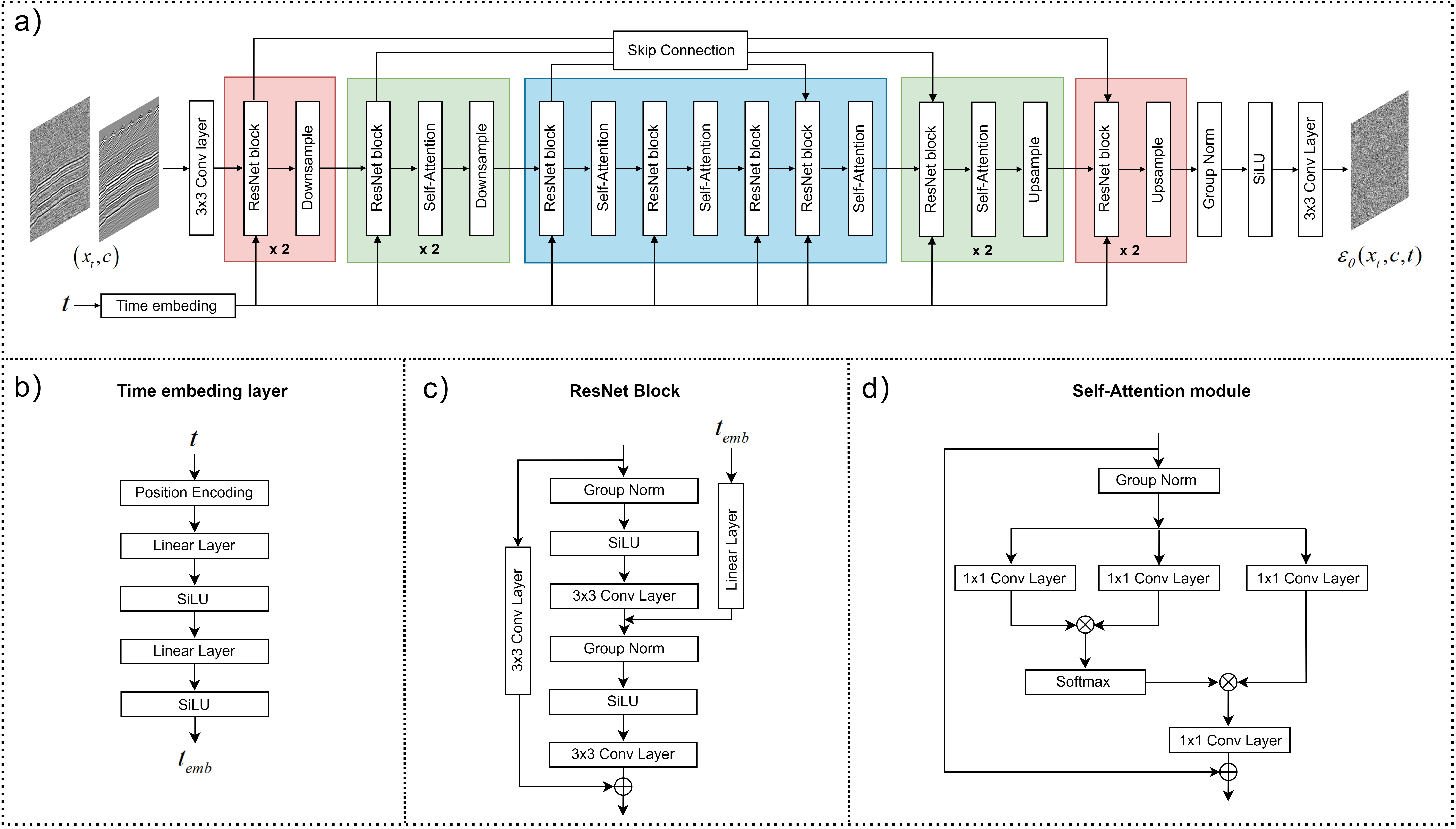}
\caption{An illustration of our network architecture. (a) The overall network structure. (b) Time embedding layer. (c) Residual block. (d) Self-attention module. In panel (a), $\times2$ indicates that the module is executed twice in sequence.}
\label{fig4}
\end{figure*}

\begin{figure*}[!t]
\centering
\includegraphics[width=6in]{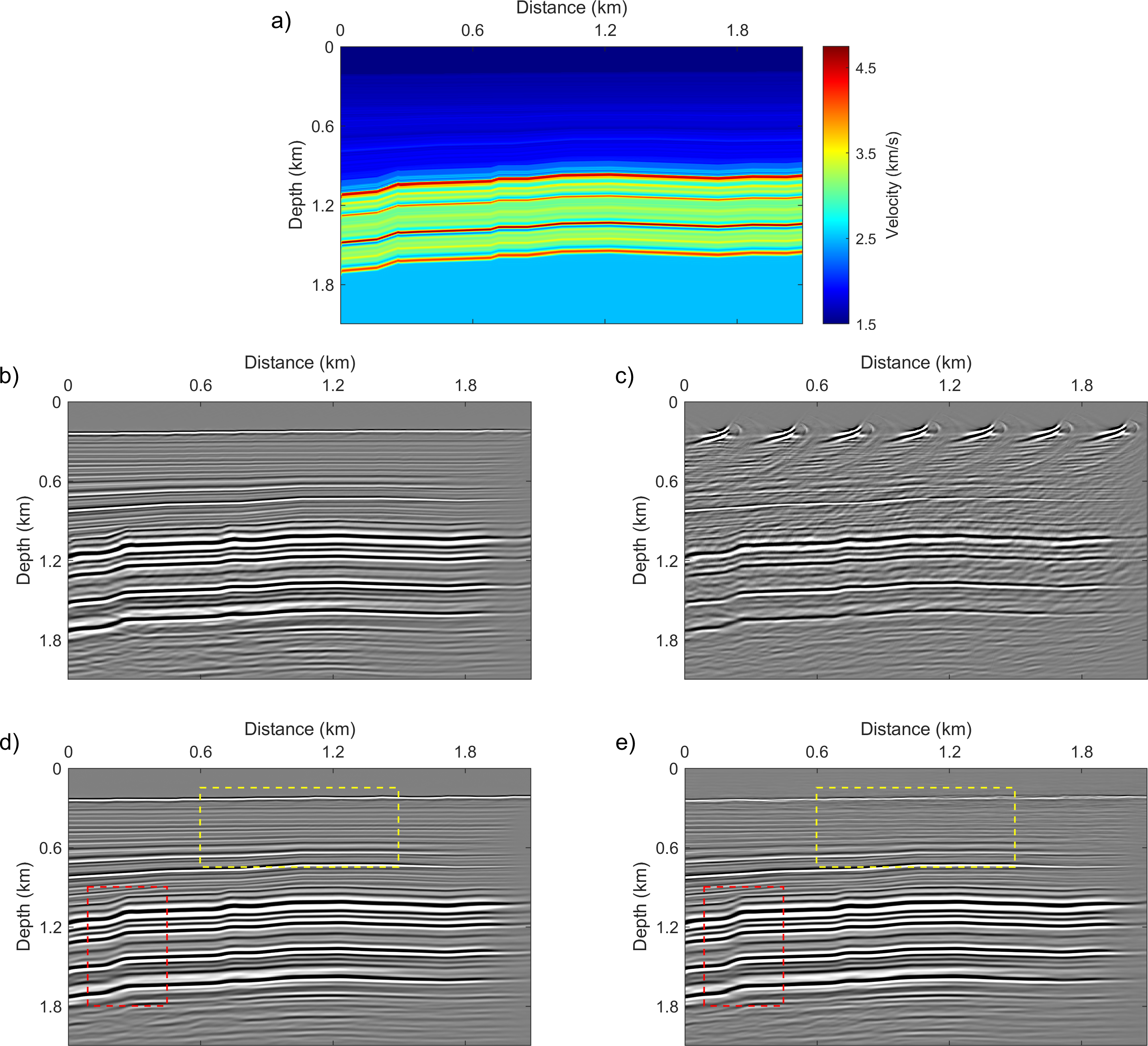}
\caption{Comparison of processing performance between our proposed method and conventional U-Net-based method for South China Sea model. (a): Velocity model corresponding to the test data. (b): Imaging result of dense data. (c): Imaging result of sparse data, used as test data. (d): Prediction result of our method. (e): U-Net's prediction result.}
\label{fig5}
\end{figure*}

\subsection{Uncertainty Quantification} 
As we introduced earlier, GDMs typically generate data by progressively denoising from pure noise. In our adaptation, even if the conditional input of the generation process (i.e., the imaging result of the sparse data) remains the same, the initial noise value and the subsequent random noise introduced at each step can lead to different generated results. This randomness is a fundamental characteristic of generative models, which allows them to produce diverse outputs. As a result, GDMs provide a natural and effective framework for uncertainty quantification. By leveraging the probabilistic nature of DDIM, we can generate multiple imaging results conditioned on the same image from sparse data and measure the variability among these results to quantify uncertainty.

Specifically, given a sparse data imaging result $I_s$ and random noise $\epsilon$ sampled from a Gaussian distribution $\mathcal{N}(0, \mathbf{I})$, the DDIM framework is used to generate a processing product $\hat{I}_d$:
\begin{equation}\label{eq7}
\hat{I}_{d} = \text{DDIM}(I_s, \epsilon).
\end{equation}

During the generation process, for each imaging result derived from sparse data, we can repeat it $B$ times along the batch size dimension. Then, we sample $B$ different noises from a Gaussian distribution. We use DDIM to process these sampled noises, using the same sparse data imaging result as the condition for all $B$ instances, thereby generating multiple prediction results. Let $\hat{I}_{d,i}$ be the $i$-th prediction, where $i\in\{1,\ldots,B\}$. These predictions are averaged to obtain the final prediction $\hat{I}_{d}$:
\begin{equation}\label{eq8}
\hat{I_d} = \frac{1}{B} \sum_{i=1}^B I_{d,i}.
\end{equation}

The uncertainty in the final prediction, denoted as $\sigma_{\hat{I_d}}$, is quantified by computing the standard deviation of the batch predictions:
\begin{equation}\label{eq9}
\sigma_{\hat{I_d}} = \sqrt{\frac{1}{B} \sum_{i=1}^B (I_{d,i} - \hat{I_d})^2}.
\end{equation}
By implementing this, we can quantify the uncertainty in the enhanced imaging results. The mean prediction $\hat{I_d}$ serves as the best estimate of the dense imaging result, while the standard deviation $\sigma_{\hat{I_d}}$ highlights areas where predictions are less reliable. As a result, it not only improves the robustness of our imaging results but also provides valuable insights into the reliability of the data processing, guiding further analysis and decision-making.

\subsection{Network Architecture} 
In this section, we illustrate the network architecture employed for enhancing sparse data imaging results. Our approach utilizes a U-Net style architecture with an encoder-decoder framework, including several key components: time embedding, residual blocks, and self-attention modules.

The general structure of the network is illustrated in Fig. \ref{fig4}(a). As stated above, the input to the network comprises a pair of images $(x, c)$, where $x_t$ represents the noisy data sampled from the dense data imaging result and \(c\) denotes the sparse data imaging result. Additionally, the network incorporates a time embedding \(t\), which is processed through a time embedding layer to integrate temporal information into the network's processing pipeline. The encoder consists of a series of residual blocks and downsampling layers, which effectively capture and compress the input features. This is followed by a central section that contains multiple residual blocks and self-attention modules, enabling the network to learn complex representations by capturing both local and global dependencies. The decoder mirrors the encoder structure, comprising upsampling layers, residual blocks, and self-attention modules. This symmetrical design ensures that high-level features learned during encoding are transformed and upsampled to reconstruct the final output $\epsilon_\theta(x, c, t)$. Skip connections between corresponding layers in the encoder and decoder enhance the network's ability to retain fine-grained details.

The time embedding layer, detailed in Fig. \ref{fig4}(b), processes the input time $t$ through sinusoidal position encoding \cite{vaswani2017attention}. This encoded time is then passed through a linear layer followed by a sigmoid linear unit (SiLU) activation function, and this sequence is repeated with another linear layer and SiLU activation to produce the final time embedding \(t_{emb}\). The time embedding is crucial for integrating temporal information into the network and is incorporated into the ResNet blocks.

Each residual block, as shown in Fig. \ref{fig4}(c), begins with group normalization \cite{wu2018group} followed by an SiLU activation and a $3\times3$ convolutional layer. This sequence is repeated and the result is added to the input through a skip connection, forming the residual block. Inclusion of time embedding $t_{emb}$ via a linear layer ensures that the temporal context influences the network's feature extraction process.

The self-attention module, illustrated in Fig. \ref{fig4}(d), starts with group normalization of the input, followed by three $1\times1$ convolutional layers. The outputs of these layers are processed through a softmax function, which computes the attention weights. These weights adjust the input feature map, enhancing the network's ability to capture long-range dependencies. More details about this module can be found in \cite{vaswani2017attention}. 

\subsection{Network Training}
Due to the lack of densely collected field data, we employ a model of the Liwan Sag region in the South China Sea to generate synthetic data for network training, utilizing the same training data as described by Cheng et al. \cite{cheng2023elastic}. The model velocities are derived from well log data. Synthetic data are generated through a finite-difference forward simulation, with a vertical grid interval of 3.0 meters and a horizontal interval of 3.1 meters. This simulation replicates 680 synthetic air gun shots at the sea surface, spaced 15.5 meters apart, using a 30 Hz Ricker wavelet. On the seabed, 3401 hydrophones are positioned to measure pressure, each spaced 3.1 meters apart. Subsequently, we extract sparse seismic data from the corresponding dense synthetic data, with an OBN spacing of 310 meters. We apply common-shot\cite{Shi2020ElasticData} and common-receiver\cite{Shi2023ElasticData} Gaussian-beam migration techniques to generate migration images for both dense and sparse node configurations. The migration images of the sparse node configuration reveal insufficient receiver illumination and gaps at shallow depths due to sparse node spacing, accompanied by migration noise. To enhance network performance, we scale the amplitudes of the imaging results to the range $(-1, 1)$ and divide the imaging results into patches of size $128\times128$. We augment the dataset by flipping the patches vertically and horizontally to increase the quantity and diversity of the training data, thus improving the generalizability of the network. In total, 16000 pairs of data patches are constructed in the dataset.

The GDM model is iteratively optimized 400,000 times using the MSE loss function, with a batch size of 32. The diffusion step $T$ for the GDM model is set to 1000. We use a cosine noise schedule, which is detailed in \cite{nichol2021improved}, to define the extent of noise at each time step. We use the AdamW optimizer \cite{loshchilov2017decoupled}, with a learning rate of 1e-4. Additionally, we train a U-Net with the same network architecture and hyperparameter settings, where the input and label for the U-Net are the imaging results from sparse and dense data, respectively, to serve as our benchmark. Training and testing were conducted on an NVIDIA A100 [80GB] graphics processing unit.
\section{Applications}

\begin{figure*}[!t]
\centering
\includegraphics[width=6in]{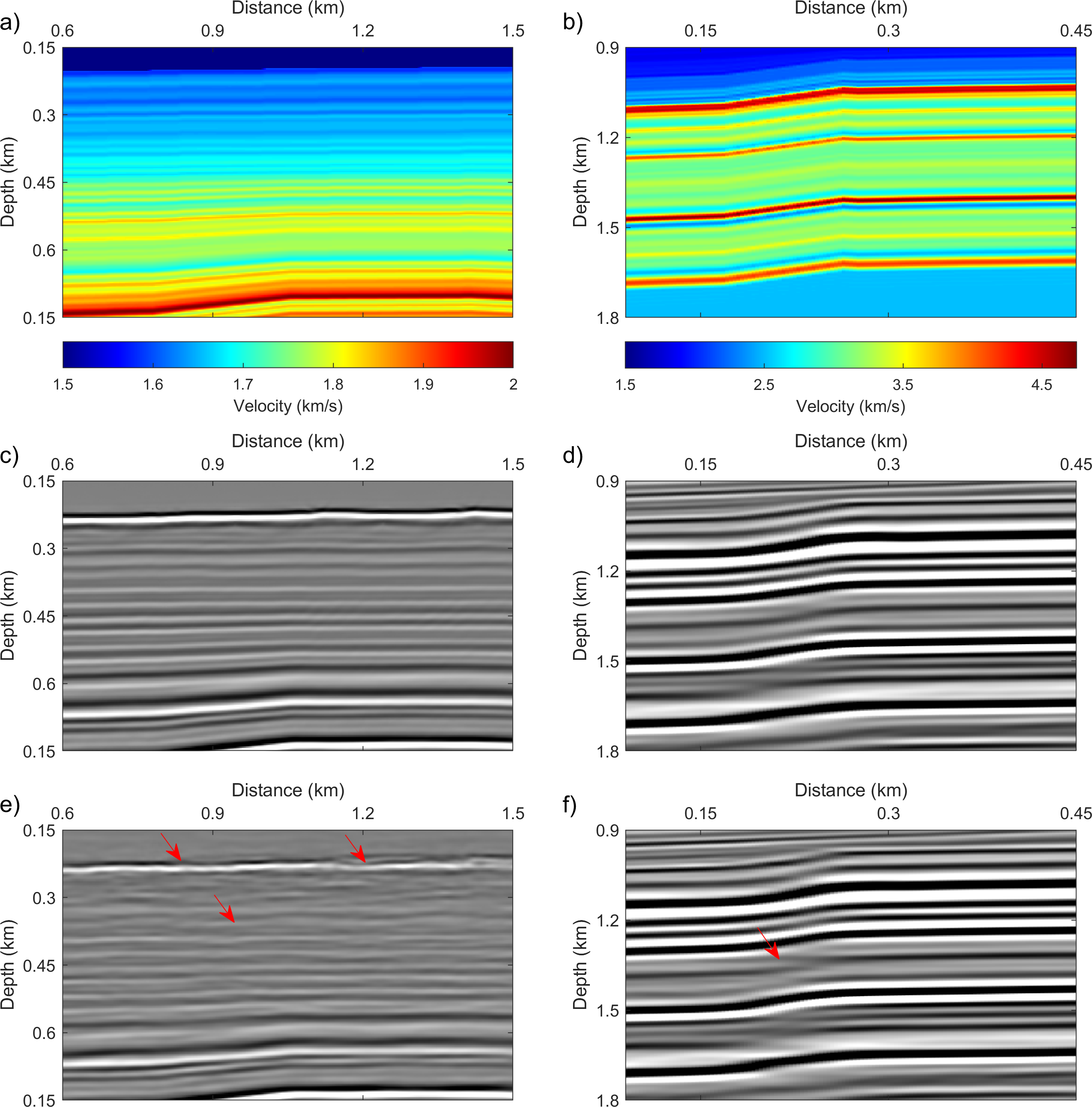}
\caption{A zoomed-in view of the yellow and red areas in Fig.\ref{fig5}. The first column corresponds to the yellow area, with the velocity model, our method’s prediction result, and the U-Net prediction result displayed from top to bottom. The second column corresponds to the red area.}
\label{fig6}
\end{figure*}

\begin{figure*}[!t]
\centering
\includegraphics[width=6in]{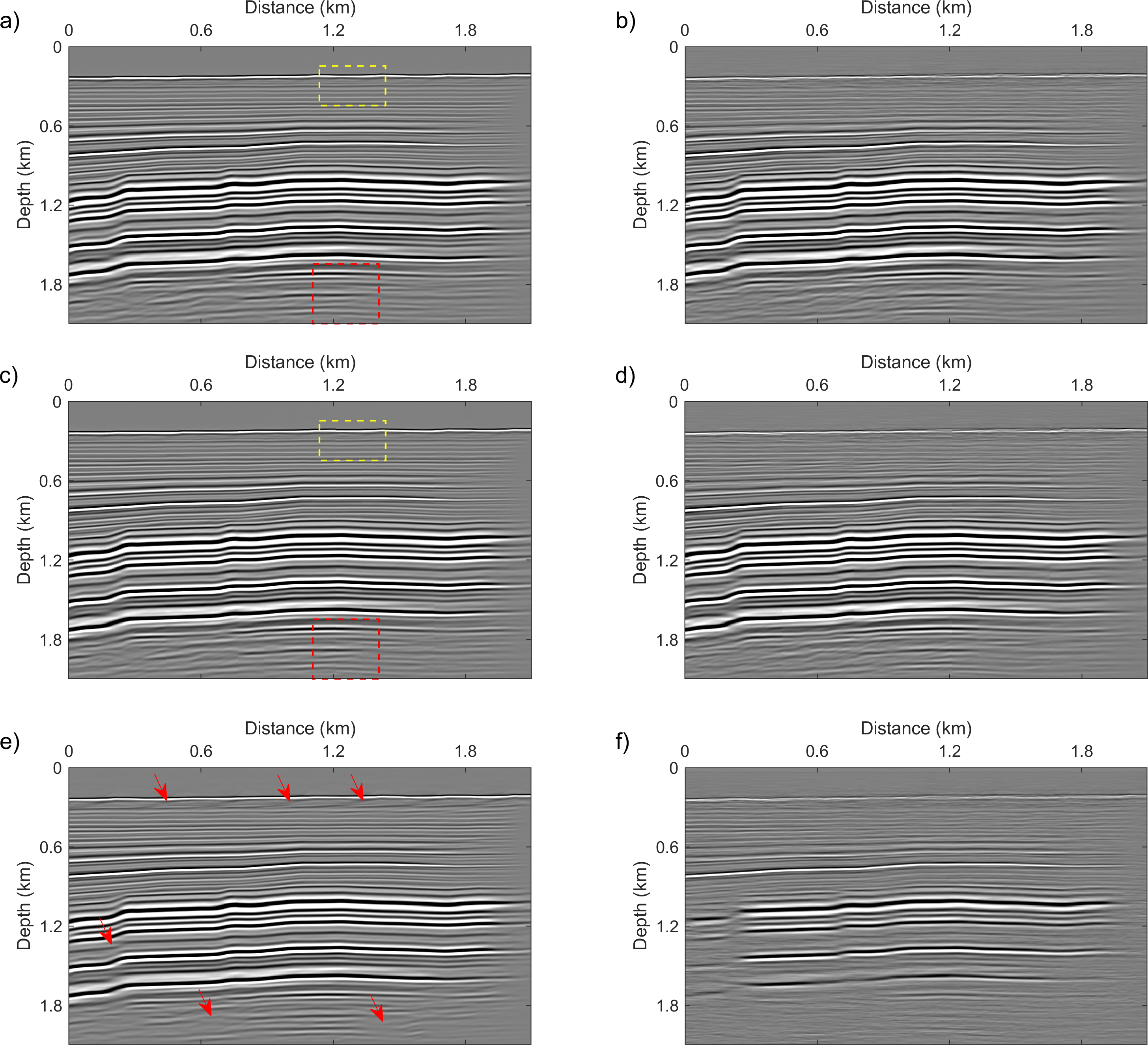}
\caption{Prediction results using different strategies for South China Sea model. The first column corresponds to the prediction results of our method and the second column corresponds to the U-Net prediction results. Each column, from top to bottom, shows the following:
The first row uses the proposed patch fusion strategy. The second row’s strategy involves simply summing the prediction results of the test data blocks and then averaging in the overlapping regions. The third row shows the prediction results for the entire test data without splitting into patches.}
\label{fig7}
\end{figure*}

\begin{figure*}[!t]
\centering
\includegraphics[width=6in]{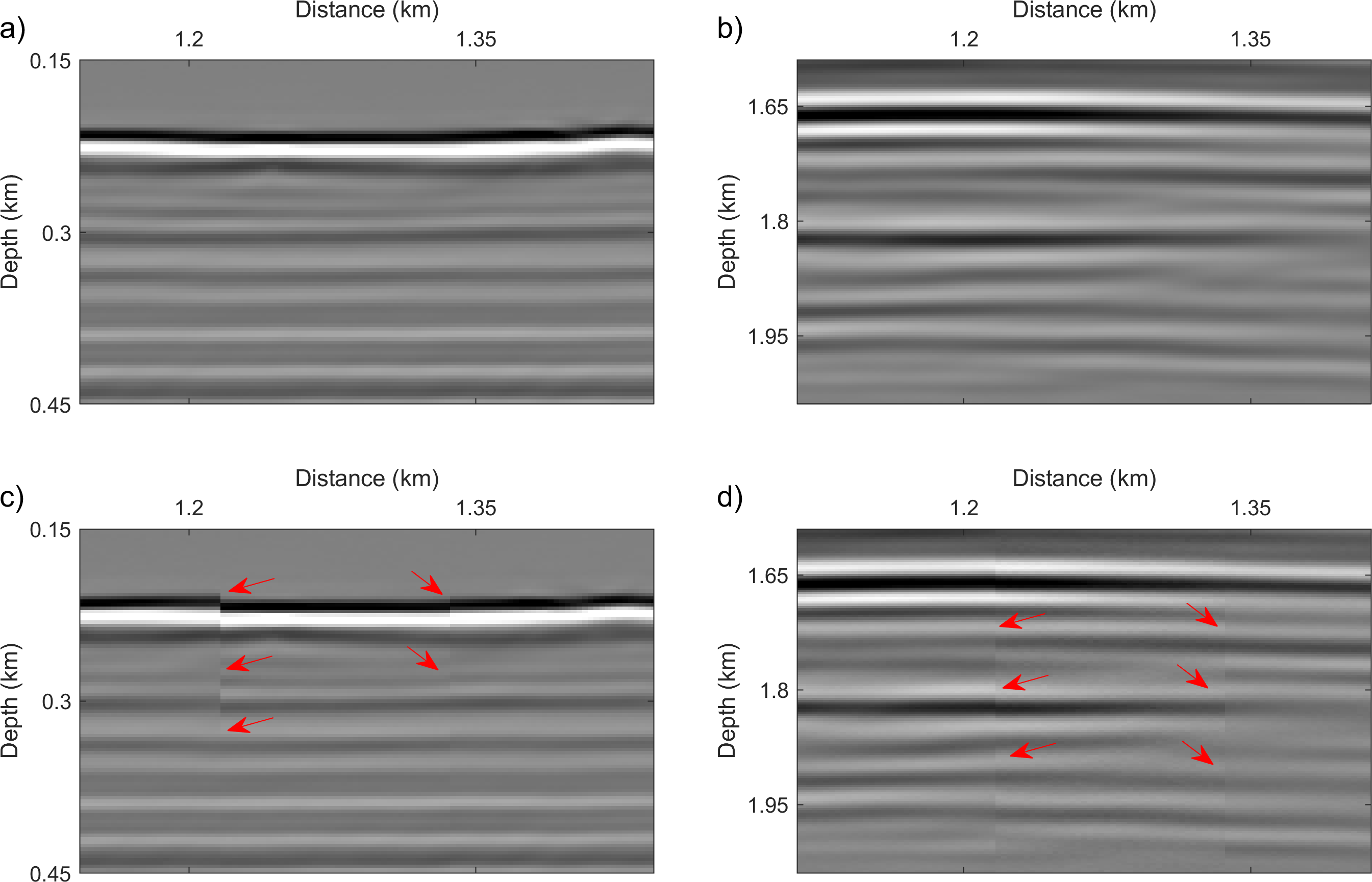}
\caption{A zoomed-in view of the red and yellow boxes in Fig. \ref{fig7}. The first and second rows represent our patch fusion results and the simple summation results, respectively. The first and second columns correspond to the yellow and red boxes, respectively.}
\label{fig8}
\end{figure*}

\begin{figure*}[!t]
\centering
\includegraphics[width=6in]{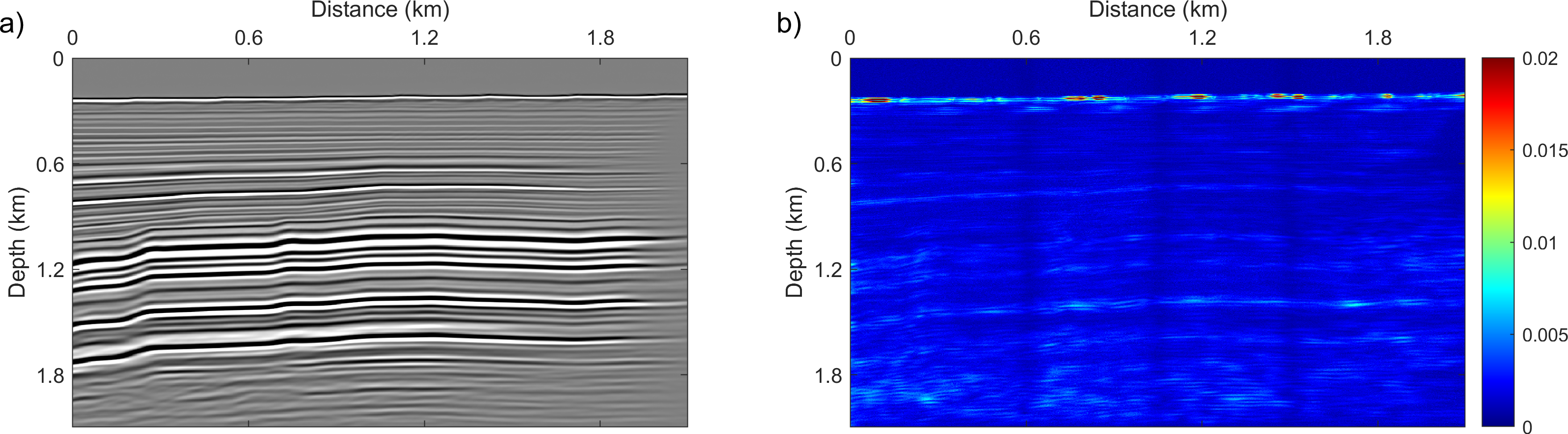}
\caption{The prediction results of our method for South China Sea model and the corresponding uncertainties.}
\label{fig9}
\end{figure*}

\begin{figure*}[!t]
\centering
\includegraphics[width=6in]{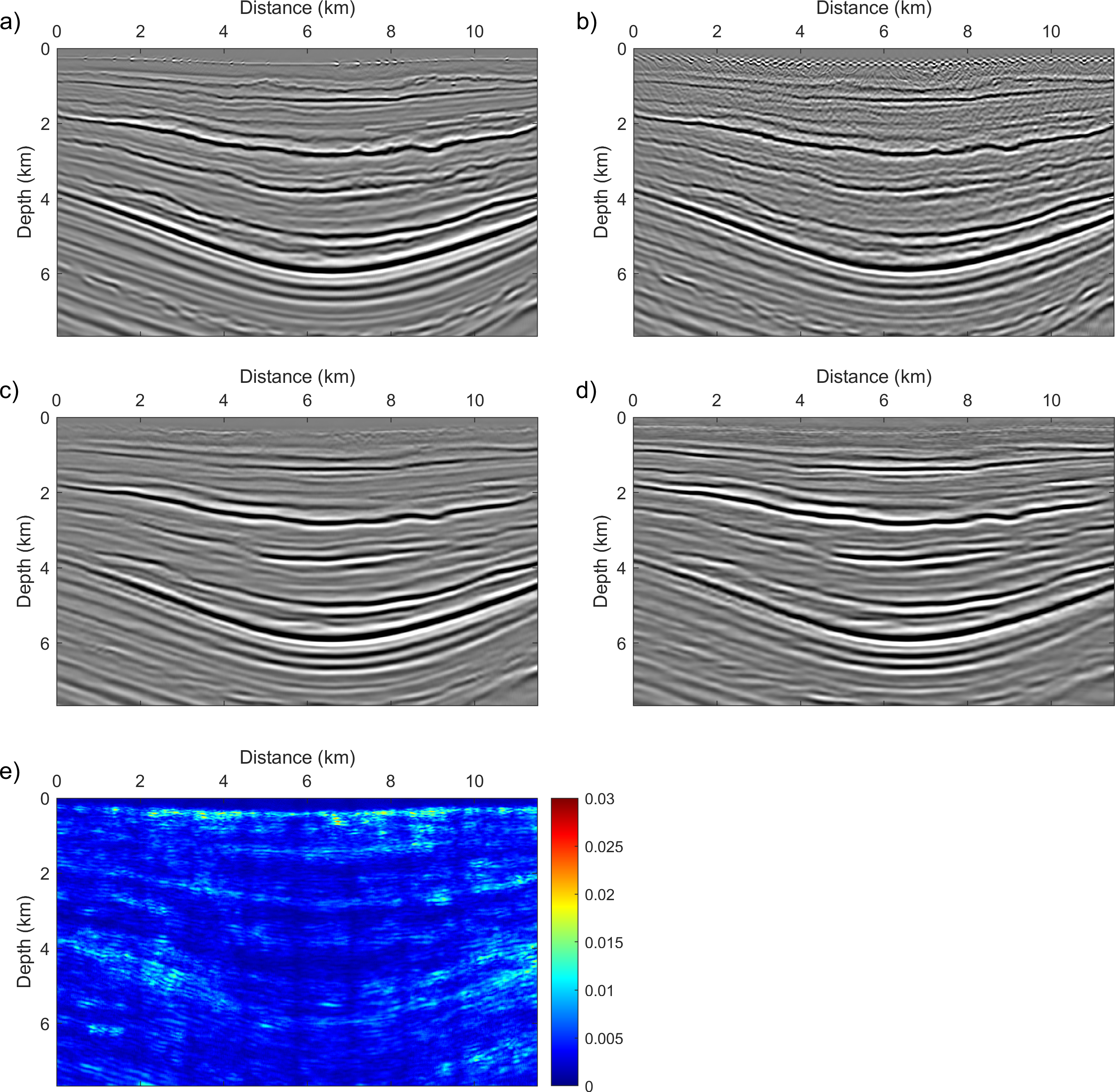}
\caption{Comparison of processing performance between our proposed method and conventional U-Net-based method for SEAM model. (a) and (b): The imaging results of dense and sparse data, respectively. (c): The prediction result of our method. (d): The U-Net's prediction result. (e) The uncertainty of our method’s prediction result.}
\label{fig10}
\end{figure*}

\begin{figure}[!t]
\centering
\includegraphics[width=3in]{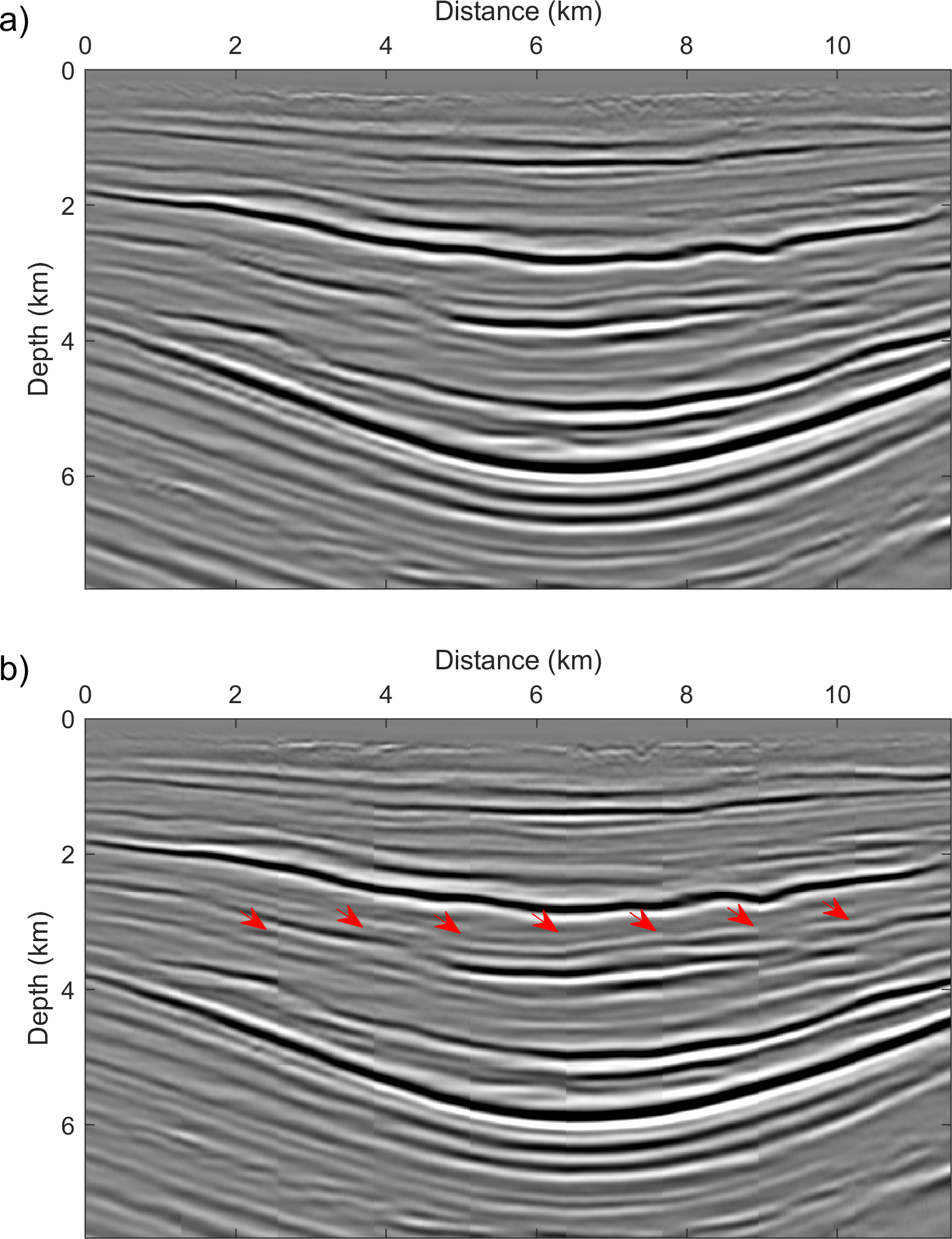}
\caption{Prediction results using different strategies for SEAM model. The first row uses the patch fusion strategy. The second row’s strategy involves simply summing the prediction results of the test data patches and then averaging in the overlapping regions.}
\label{fig11}
\end{figure}

\begin{figure*}[!t]
\centering
\includegraphics[width=6in]{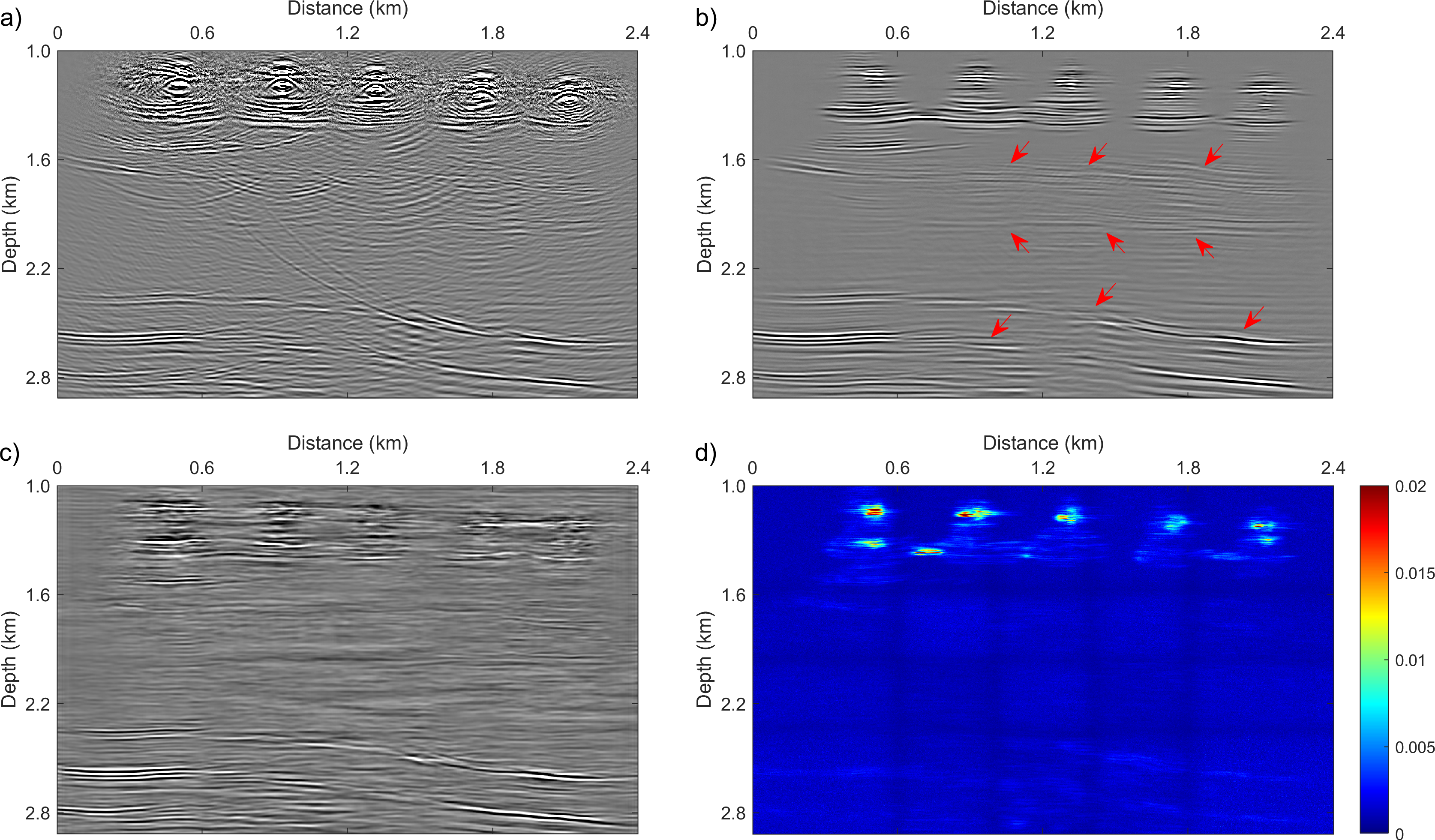}
\caption{Comparison of processing performance between our proposed method and conventional U-Net-based method for sparse OBN field data acquired from South China Sea. (a): The migration result from original sparse data. (b): The predicted result of our method. (c): The U-Net's prediction result. (d): The uncertainty of our prediction result.}
\label{fig12}
\end{figure*}

In the following, to validate the effectiveness of our method, we will share our test results on synthetic and field data. We emphasize that during the generation process, we used the DDIM model to resample the reverse process, aiming to accelerate inference. Specifically, we resample the original 1000 diffusion steps to 50 steps, which means that the generation process skips 20 (i.e., $1000/50$) steps at each time. When quantifying the uncertainty of the processed results, we set the batch size dimension to 10, which means $B$ in equation \ref{eq8} is defined as 10. When performing inference using the patch fusion strategy, we set the patch size (i.e., $p\times p$) to $256\times256$. For the step size $s$, the South China Sea model is set to 148, while the SEAM model and field data are set to 128.

\subsection{South China Sea Model}
We first conduct tests on an additional section of the South China Sea model. Fig. \ref{fig5}(a) illustrates the velocity model corresponding to the test data. The model consists of a water layer, multiple thin and flat transition layers designed to replicate the ‘soft water bottom’ effect, and a sequence of sedimentary rock layers. The grid spacing for the model is 3.0 meters vertically and 3.1 meters horizontally. Synthetic air gun shots were fired at the sea surface at 15.5-meter intervals, and a 30 Hz Ricker wavelet was employed in the forward modeling. 

The imaging results from dense data with a group interval of 3.1 meters [Fig. \ref{fig5}(b)] and sparse data with a group interval of 310 meters [Fig. \ref{fig5}(c)] are presented, along with the prediction results of our method [Fig. \ref{fig5} (d)] and the conventional U-Net model [Fig. \ref{fig5}(e)]. Fig. \ref{fig5}(c) illustrates that the image obtained from sparse acquisition exhibits discontinuities and gaps compared to that obtained from dense acquisition and is significantly affected by noise. Particularly in the shallow section, the imaging quality is poorer due to the narrow illumination cones of the seismic waves at shallow depths. In contrast, after processing these images with the trained networks, the resulting images, as shown in Figs. \ref{fig5}(d) and \ref{fig5}(e), demonstrate substantial improvements in both continuity and noise reduction. To further compare the performance between our method and the conventional U-Net, Fig. \ref{fig6} provides an enlarged comparison of the specific areas highlighted in Fig. \ref{fig5}. In Fig. \ref{fig6}, the first and second columns correspond to the yellow and red boxes, respectively. Each column, from top to bottom, displays the velocity model, our method's prediction results, and conventional U-Net's prediction results. We can see that our method, based on the GDM, also uses the U-Net architecture as our baseline but demonstrates superior performance compared to conventional U-Net-based methods. This demonstrates the superior capability of the generative model compared to conventional CNN-based methods. This ability is attributed to its powerful ability to learn and extract complex data distribution features from given training data.

Fig. \ref{fig7} compares the prediction results obtained using different inference strategies: 1. The proposed patch fusion strategy; 2. Dividing the test data dividing the test data into small overlapping patches, predicting, and then directly averaging the overlapping areas; 3. Performing direct prediction on test data without splitting into patches. In Fig. \ref{fig7}, the first and second columns correspond to the prediction results of our method and the conventional U-Net, respectively. Each column, from top to bottom, corresponds to the first, second, and third inference strategies. We can see that for both our method and conventional U-Net, predicting on patches and then combining them (i.e., strategies 1 and 2) significantly improves prediction accuracy compared to directly predicting on the test data (i.e., strategy 3). For our method, if we directly predict the entire test data, it results in decreased performance in handling shallow, deep, and areas with significant lateral structural changes. For the U-Net method, the difference between patch prediction and direct prediction strategies is substantial, with the direct prediction strategy leading to a very poor prediction product.

To further compare our proposed strategy 1 with strategy 2, we show the zoomed-in views of the yellow and red boxes in Fig. \ref{fig8}. The first and second columns correspond to strategies 1 and 2, respectively, while the first and second rows correspond to the yellow and red boxes, respectively. We can observe that in both regions, the processing results from our proposed strategy show better continuity, whereas the prediction results from strategy 2 exhibit noticeable discontinuities and artifacts due to the simple averaging.

Fig. \ref{fig9} presents the prediction results of our method along with the corresponding uncertainties, providing a measure of confidence in the predictions. It should be noted that the U-Net prediction results cannot offer uncertainty estimates, which highlights the advantage of our method. We can observe that the shallow regions where OBNs are located exhibits high uncertainty. This is because the large spacing between nodes results in insufficient illumination of shallow reflection points, which results in poor continuity of shallow reflection events. This poses a challenge for prediction using the trained GDM. As a result, the confidence in the prediction output in this area is relatively low.

\subsection{SEAM Model}
Next, we evaluate the SEAM model that replicates the complex geological formations of the deep-water region in the Gulf of Mexico, to further validate the effectiveness and generalization ability of our approach. 

Fig. \ref{fig10} presents the prediction results for this model. Specifically, Figs. \ref{fig10}(a) and \ref{fig10}(b) display the images derived from dense and sparse data, respectively, which provide a clear baseline for comparison. Fig. \ref{fig10}(c) shows the prediction results of our method, while Fig. \ref{fig10}(d) presents the U-Net prediction results. Furthermore, Fig. \ref{fig10}(e) illustrates the uncertainty associated with the predictions of our method, providing a measure of confidence in the results. We can see that the prediction result of our method highlight the effectiveness of our approach, particularly compared to the U-Net's prediction result. Meanwhile, this comparison suggests that the GDM has better generalization ability compared to the conventional CNN-based methods. Similar to what we observed in the previous synthetic test, the processing result near the shallow water layer here show greater uncertainty. This is because the sparse data acquisition leads to limited illumination in this area, causing significant discontinuities in the imaging results, which in turn result in higher uncertainty in the processing result. Also, in areas with steep dips and fold structures, the uncertainty in the prediction result increases. This is expected due to the complex geological structures in these regions.

Fig. \ref{fig11} compares the prediction results obtained using different inference strategies, where panels (a) and (b) represent the predicted results using strategies 1 and 2 (mentioned in the previous synthetic data test), respectively. Here, strategy 3 is not applicable because the size of test data (i.e., $832\times1216$) is too large. When we directly predict on the test data, it exceeds the GPU memory (80GB), which also demonstrates that using the patch strategy can reduce the memory burden. In this figure, the first and second rows represent the prediction result using strategies 1 and 2, respectively. We can see that the proposed patch fusion strategy contributes to a more continuous imaging improvement result. In contrast, the second strategy leads to a poor continuity and noticeable discontinuity effect caused by splitting patches, as indicated by red arrows. This further underscores the importance of the patch fusion strategy.

\subsection{Field Data}
Finally, we evaluate our method using real seismic data to further validate its effectiveness and robustness. These field data were acquired from the Liwan Sag region in the South China Sea, using only 5 OBNs with an approximate spatial interval of 400 meters along the horizontal direction.

Fig. \ref{fig12} presents the processing results. Specifically, panel (a) corresponds to the migration result from the sparse OBN data, which serves as a baseline for comparison. Panels (b) and (c) show the predicted result of our method and the U-Net-based method, respectively. We can see that the image produced by our method exhibits enhanced continuity and reduced noise, indicating a significant improvement over the original migration result. Especially in shallow areas, our method effectively reveals two events. Although the U-Net-based method also removes some arc-shaped noise caused by sparse acquisition, it introduces very strong artifacts in the entire image. This comparison once again demonstrates the superior performance of our method compared to the U-Net-based method.

Fig. \ref{fig12}(d) illustrates the uncertainty associated with our prediction result. It also shows high uncertainty in shallow layers close to seawater. However, the two layers revealed in the shallow section by our method show very low uncertainty. This indicates that these two layers are highly likely to be real and actually present. 

% This uncertainty map provides a measure of confidence in the predictions made by our method. Areas with higher uncertainty are clearly identified, allowing for a more nuanced interpretation of the seismic data. The ability to quantify uncertainty is a distinct advantage of our method, as it provides additional insights into the reliability of the predictions.

% However, the U-Net-based method 

% Panel (b) shows the predicted result of our method, demonstrating its effectiveness in processing field data. The image produced by our method exhibits enhanced continuity and reduced noise, indicating a significant improvement over traditional methods. This result underscores the capability of our approach to handle real-world seismic data with high accuracy. Panel (c) presents the prediction result obtained using the U-Net model, serving as a comparative benchmark. The U-Net model, while effective, does not achieve the same level of detail and noise reduction as our method. This comparison highlights the superior performance of our approach in terms of image clarity and structural continuity. 

% This migration result is derived from conventional seismic processing techniques and provides a reference point for assessing the performance of our method. 

\section{Discussion}
In this section, we will first explain how our method generates an improved imaging result. Then, we will analyze how the settings of some key components affect the performance of our method. Finally, we will discuss some current limitations of this method and outline future research directions.

\subsection{The Generation Process}
In the previous tests, we demonstrated that the proposed method significantly improves the imaging results derived from sparse acquisition for both synthetic and field data. Our method is adapted from generative diffusion models (GDMs). Thus, how does it generate an enhanced imaging result from randomly sampled noise, guided by the sparse data imaging result, step by step?

\begin{figure*}[!t]
\centering
\includegraphics[width=6in]{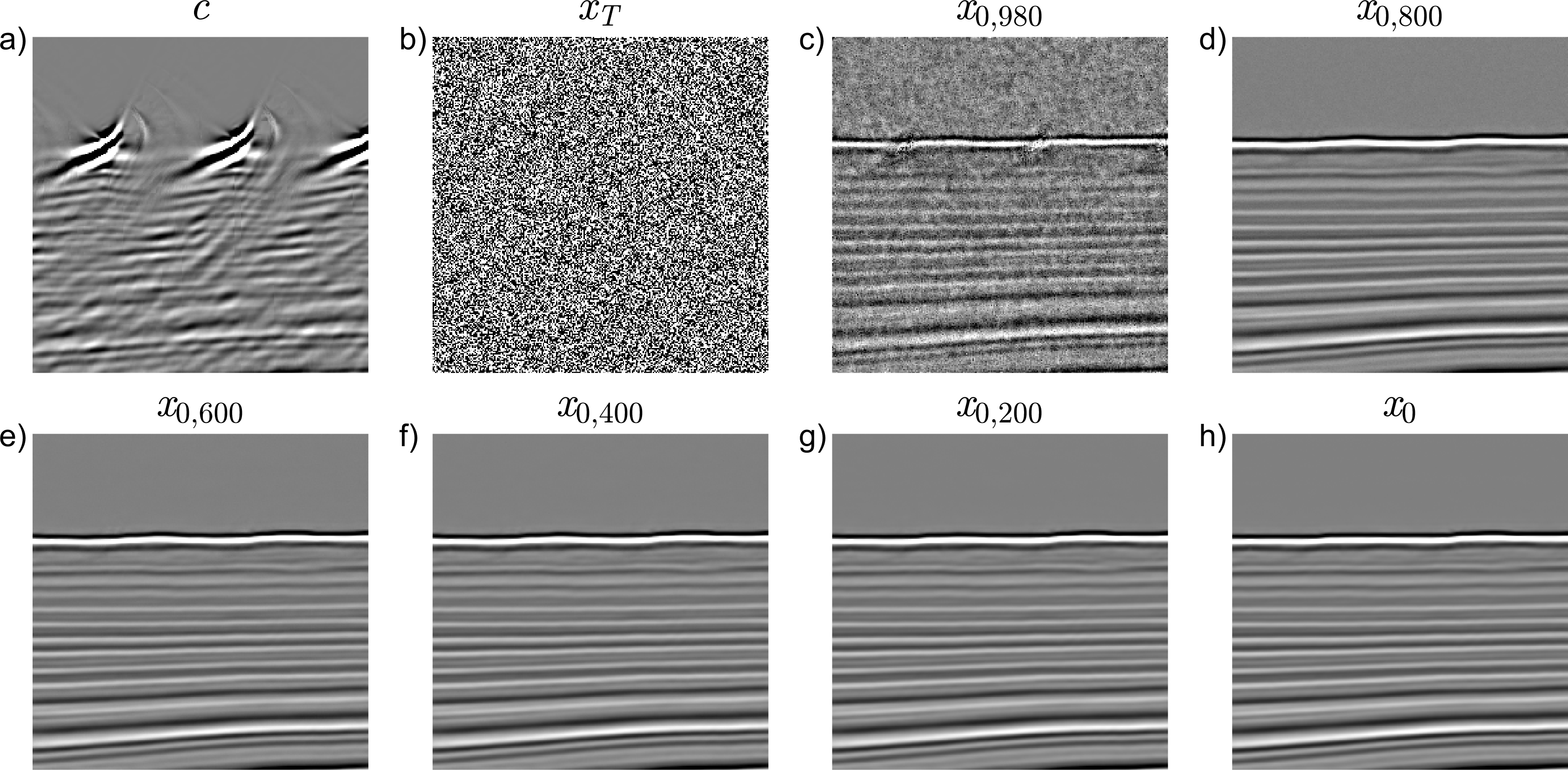}
\caption{The network’s prediction of $x_0$ at different sampling time steps. (a): The image result for sparse data (b): The sampled random noise. (c)-(g): The network’s prediction of $x_0$ at different sampling steps, e.g., 980, 800, 600, 400, and 200. (h): The final prediction result.}
\label{fig13}
\end{figure*}

To illustrate this, we present such a generation process in Fig. \ref{fig13}. Panel (a) shows the imaging result from sparse data, and panel (b) represents the randomly sampled noise, which together form the initial input to the network. Panels (c) to (g) show the estimated enhanced imaging results at 980, 800, 600, 400, and 200 sampling steps, respectively, while panel (h) shows the final predicted result. We emphasize that here we used DDIM to resample the 1000 sampling steps to 50 steps. Therefore, the 980, 800, 600, 400, and 200 sampling steps correspond to 49, 40, 30, 20, and 10 sampling steps in the DDIM model, respectively. We can see that in the initial sampling steps, such as at 980 steps, the network's estimated results contain a lot of noise. As the sampling process progresses, our method gradually removes noise, improves continuity, and reduces artifacts. After completing all sampling steps, our method provides a reliable predicted result. This trend underscores the ability of GDMs to refine their predictions over time, ultimately leading to high-quality results.

\subsection{The Impact of Different Sampling Steps}
To reduce the computational burden caused by the sampling process, we used the DDIM model, which allows the resampling of time steps, offering a significant advantage over the DDPM by reducing the time required for the generation process. Does this resampling lead to a decrease in the quality of the predicted results? Additionally, do the generated results differ significantly with different sampling steps?

\begin{figure*}[!t]
\centering
\includegraphics[width=6in]{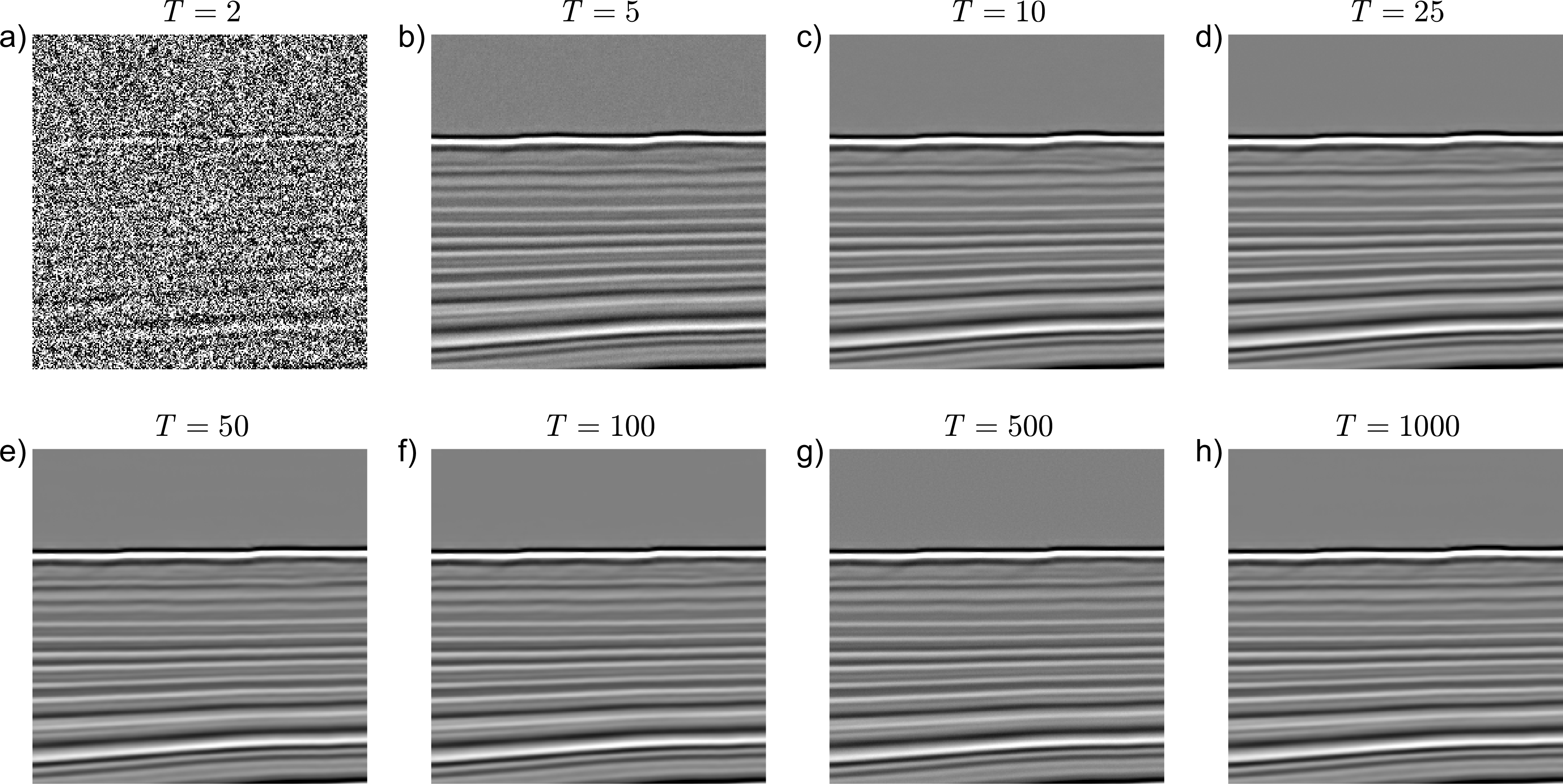}
\caption{The generation results using different sampling steps. (a) through (h) illustrate the generated results using 2, 5, 10, 25, 50, 100, 500, and 1000 time steps, respectively. }
\label{fig14}
\end{figure*}

To answer these questions, we here test different resampling strategies, specifically examining the impact of different sampling steps on the quality of the generated results. Fig. \ref{fig14} presents the generation results using different sampling steps. Panels (a) through (h) illustrate the results for the 2, 5, 10, 25, 50, 100, 500, and 1000 time steps, respectively. In smaller time steps, such as 2, noticeable noise is present in the generated result. By the 5 steps, the noise is visibly reduced, indicating an improvement in the quality of the generated result. As the number of sampling steps increases, it is evident that the quality of the generated results gradually improves. However, we should note that the time required for the generation process also increases almost linearly with the number of sampling steps. To balance the trade-off between the time cost and the quality of the generated results, we selected 50 time steps for our numerical experiments. This choice provides a reasonable compromise that ensures high-quality results while maintaining a manageable generation time.

\subsection{Uncertainty Quantification Using Different Batch Size}
As stated above, to quantify the uncertainty of our processing result, we duplicated the imaging results of sparse data along the batch size dimension by $B$ times and similarly sampled $B$ random noises. These were combined as initial input for the sampling process. We averaged the sampling results along the batch dimension to obtain our final prediction. The standard deviation of these sampled results was used as our measure of uncertainty. So, does the different replication factor, i.e., the number of $B$, affect our prediction results and uncertainty quantification?

\begin{figure*}[!t]
\centering
\includegraphics[width=6in]{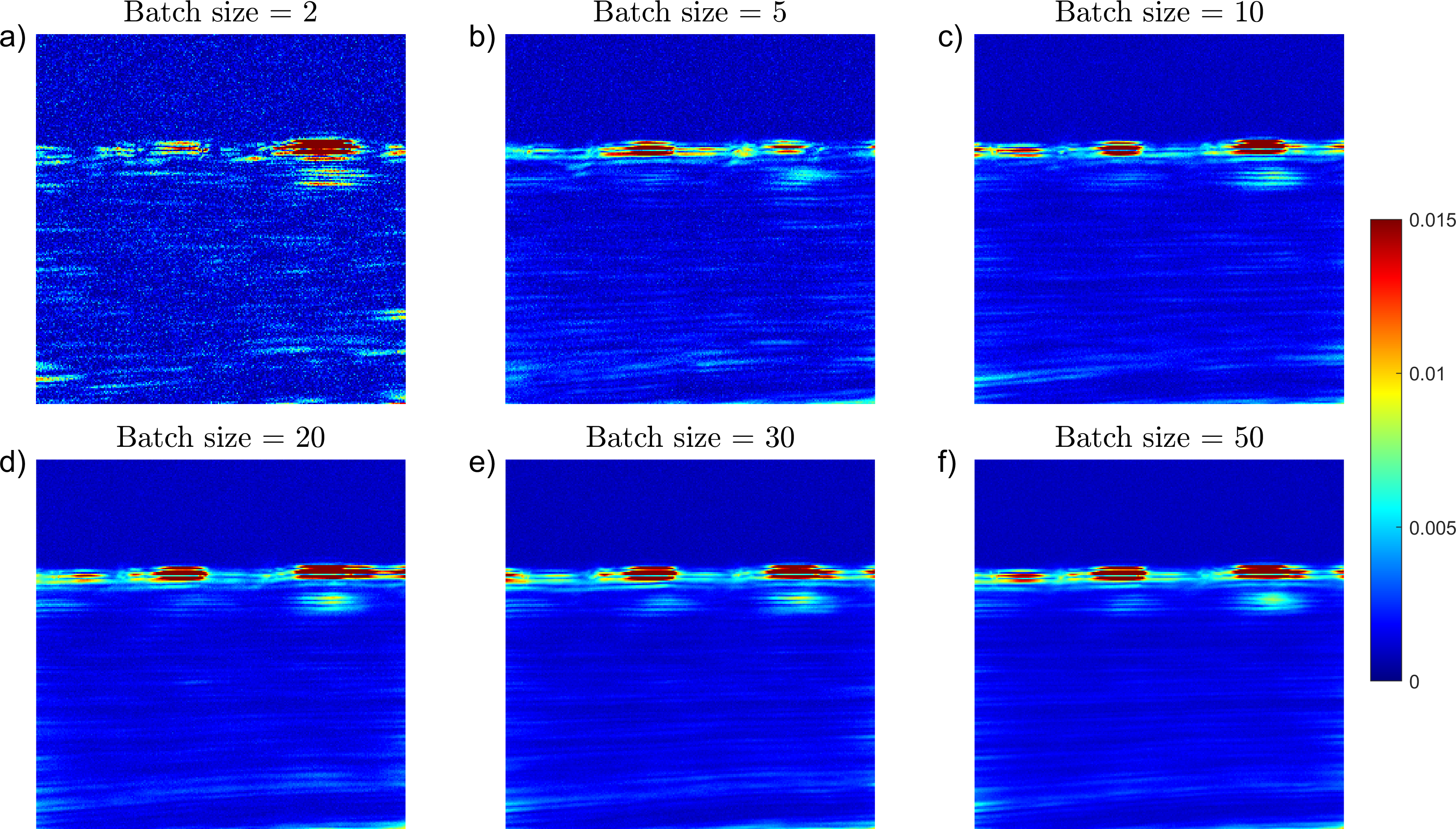}
\caption{Uncertainty maps for different batch sizes: (a) Batch size = 2, (b) Batch size = 5, (c) Batch size = 10, (d) Batch size = 20, (e) Batch size = 30, (f) Batch size = 50.}
\label{fig15}
\end{figure*}

\begin{figure*}[!t]
\centering
\includegraphics[width=6in]{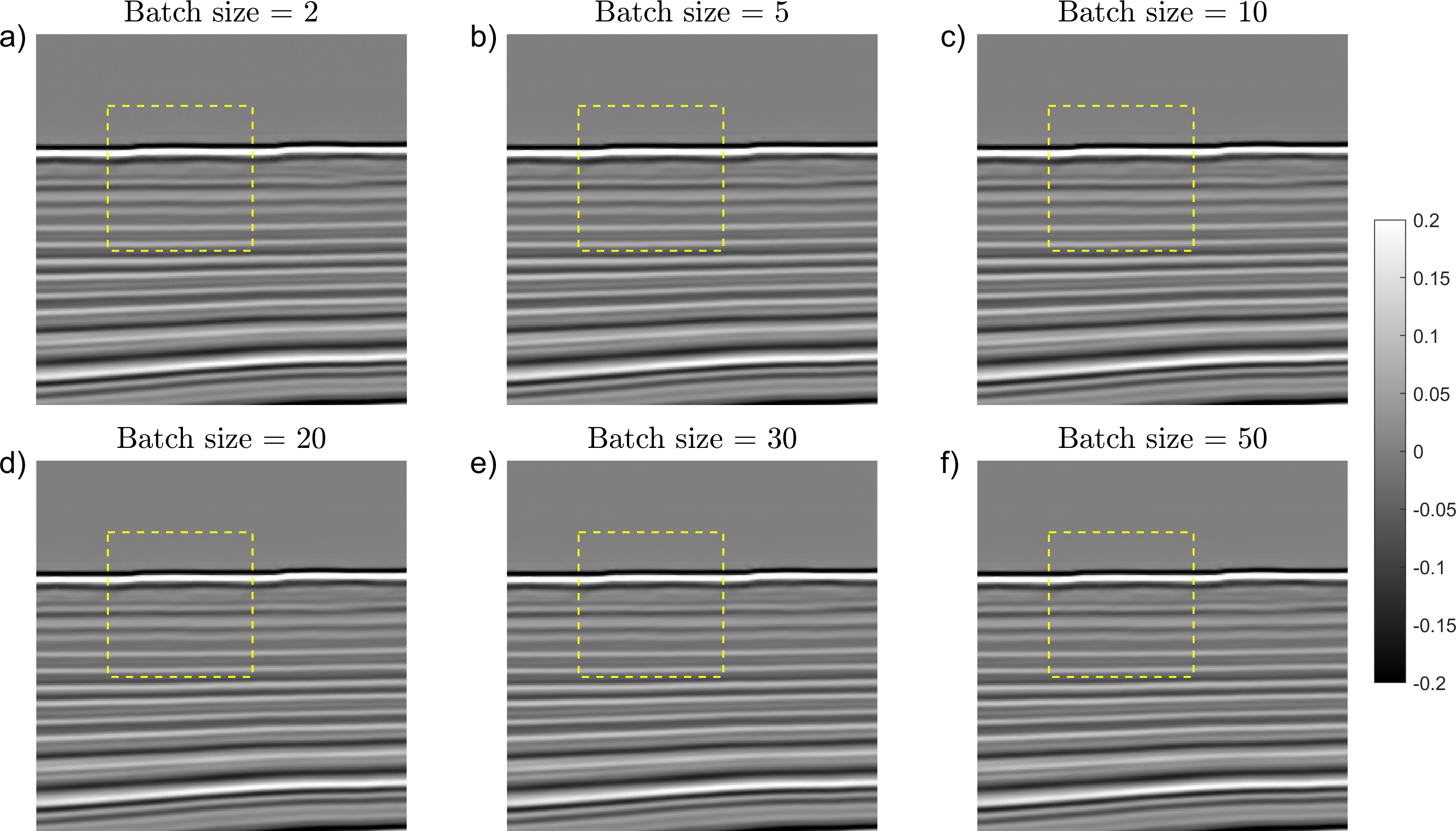}
\caption{Mean of the sampled results along the batch dimension, serving as our prediction.}
\label{fig16}
\end{figure*}

\begin{figure*}[!t]
\centering
\includegraphics[width=6in]{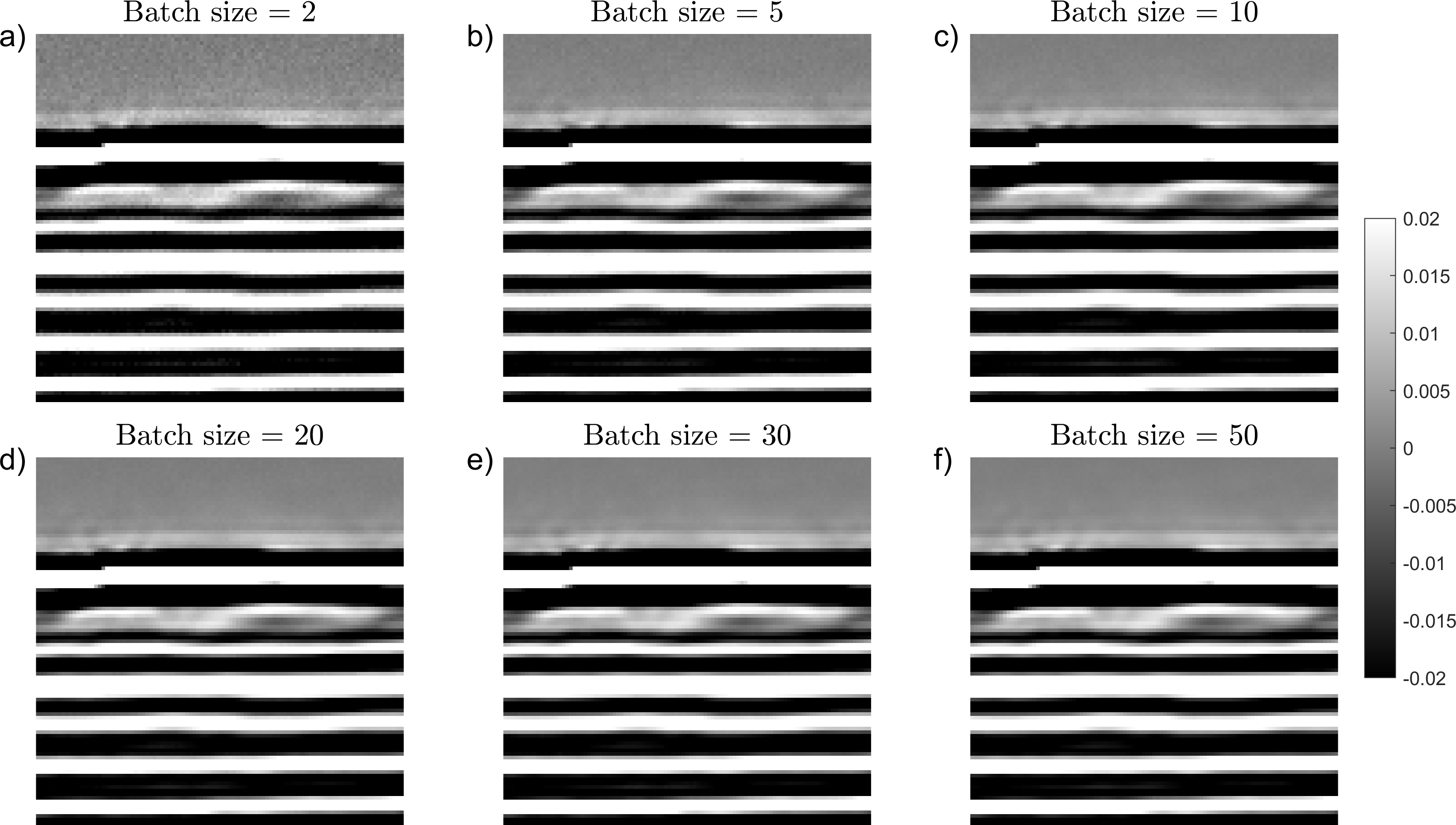}
\caption{A zoomed-in view of the yellow box in Fig. \ref{fig16}, with the clip magnified 10 times.}
\label{fig17}
\end{figure*}

Consequently, we analyze the impact of batch size $B$ on uncertainty quantification. Figs.\ref{fig15}(a) to (f) illustrate the uncertainty for batch sizes of 2, 5, 10, 20, 30, and 50, respectively. We can see that the contamination in the uncertainty maps decreases as the batch size increases. To further analyze this, Fig. \ref{fig16} presents the mean of the sampled results, which serves as our prediction. The yellow box in Fig. \ref{fig16} is enlarged and displayed in Fig. \ref{fig17}, with the clip magnified 10 times. We can observe that a larger batch size results in a more stable prediction with less noise, making it more reliable. However, a larger batch size also leads to increased computational and memory costs. Table \ref{tab1} lists the inference time and memory consumption for this analysis. Therefore, to balance these factors in our numerical experiments, we selected a batch size of 10.

\begin{table}
\centering
\caption{The comparison of computation and memory cost using different batch size.}
\renewcommand\arraystretch{1.5}
\begin{tabular}{ccc}
    \hline
    \text {Batch size} & \text { Inference time (s) }  & \text { Memory cost (GB) } \\
    \hline
    $2$ & $4.448$ & $4.88$ \\
    $5$ & $9.845$ & $8.08$ \\
    $10$ & $19.028$ & $12.74$ \\
    $20$ & $37.519$ & $22.27$ \\
    $30$ & $56.710$ & $31.81$ \\
    $50$ & $89.697$ & $50.91$  \\
    \hline
\end{tabular}
\label{tab1}
\end{table}

\subsection{Computational and Memory Consumption Caused By Patch Fusion Strategy}
As we demonstrated earlier, introducing the patch fusion strategy not only effectively reduces memory burden but also significantly enhances the quality of the generated products. However, does it lead to significant computational cost compared to directly predicting the entire data, and how substantial is the reduction in memory consumption?

To illustrate this, Table \ref{tab2} presents the time and memory consumption during inference using the patch fusion strategy and the direct prediction strategy on the synthetic South China Sea model. It is evident that even when splitting the data into multiple patches for prediction, this does not increase the inference time. On the contrary, it takes less time compared to directly predicting the entire dataset. Moreover, we can see that the patch fusion strategy reduces memory consumption by an order of magnitude. This reduction is crucial for practical applications, as seismic imaging results in real scenarios typically have very large sizes.

\begin{table}
\centering
\caption{The comparison of computation and memory cost using different inference strategies.}
\renewcommand\arraystretch{1.5}
\begin{tabular}{ccc}
    \hline
    \text {Strategy} & \text { Inference time (s) }  & \text { Memory cost (GB) } \\
    \hline
    \text{With Patch Fusion} & $47.658$ & $4.02$ \\
    \text{Without Patch Fusion} & $56.357$ & $47.49$ \\
    \hline
\end{tabular}
\label{tab2}
\end{table}

\subsection{Limitations and Future Work}
Since GDMs are trained on a specific data distribution and subsequently generate data within that feature range, they may suffer from generalization issues, especially when the distribution of the prediction data deviates significantly from the training data. Consequently, in the test results on the SEAM model, we observed a slight performance decline due to the minor differences in characteristics between it and our training data. In the case of field data, our trained data are derived from the model that is constructed using the well velocity from the data acquisition area, which can enhance the generalization of the trained model to some extent. Therefore, a primary focus of our future work is to improve the generalization capabilities of GDMs, as this is a prerequisite for their broader application. 
\section{Conclusion}
In this paper, we focused on the challenges associated with subsurface imaging from low-fold seismic reflection data and data with restricted acquisition geometry, such as those obtained from ocean bottom nodes. These data often suffer from discontinuities, migration swing artifacts, and inadequate resolution, which hinder accurate seismic interpretation. To address these limitations, we introduced a novel seismic resolution enhancement method based on generative diffusion models (GDMs). Compared to conventional convolutional neural network-based seismic processing solutions, our proposed method has two unique advantages: (1) GDMs offer superior capability to learn and capture the complex distribution features within seismic data; (2) GDMs provide a natural way to measure the uncertainty of our prediction results, thereby offering guidance for the decision-making process. We also proposed a patch fusion strategy to address the memory burden of GDMs and the performance degradation encountered for large-scale imaging results. The effectiveness of the proposed method was validated through experiments on both synthetic and field data. The results showed significant improvements in event continuity and noise reduction in migration products derived from sparse data acquisition. Meanwhile, the uncertainty of the processing results was provided to help us quantify the confidence and reliability in our results.

\section*{Acknowledgments}
We are grateful to the authors of the PyTorch implementation of Improved Denoising Diffusion Probabilistic Models \cite{nichol2021improved}, who we based their original code to adapt our sparse data imaging enhancement task. We also thank the Tongji University for graciously supplying testing data. This work utilized the resources of the Supercomputing Laboratory at King Abdullah University of Science and Technology (KAUST) in Thuwal, Saudi Arabia.

\bibliographystyle{IEEEtran}
\normalem    % Ignore “\usepackage{ulem}”
\bibliography{reference}{}

\end{document}